\documentclass{aa}

\usepackage{graphicx}
	\graphicspath{{./img/},{./}}
\usepackage{txfonts}
\usepackage{hyperref}
\usepackage{subcaption}
\usepackage{xcolor}
\usepackage{textcomp,mathabx}
\usepackage{siunitx}
    \DeclareSIUnit{\astronomicalunit}{AU}
	\DeclareSIUnit{\parsec}{pc}
	\DeclareSIUnit{\earthmass}{M_\Earth}
	\DeclareSIUnit{\solarmass}{M_\Sun}
	\DeclareSIUnit{\jupitermass}{M_J}
	\DeclareSIUnit{\year}{yr}
    \DeclareSIUnit{\jansky}{Jy}
    \DeclareSIUnit{\arcsecond}{arcsec}
\usepackage{placeins}

\begin{document} 

    \bibliographystyle{aa}
    \title{Emergence of streamers in simulations of late infall}

    \author{L.-A. H\"uhn
        \and
        C. P. Dullemond
        }

    \institute{Institut für Theoretische Astrophysik, Zentrum für Astronomie der Universität Heidelberg, Albert-Ueberle-Str. 2, 69120 Heidelberg, Germany\\
            \email{huehn@uni-heidelberg.de}
            }

    \date{\today}
 
    \abstract
    {Growing observational evidence suggests that Class II protoplanetary disks may undergo substantial interactions with their environment in the form of late infall. This mass inflow predominantly manifests itself in the form of so-called streamers: filaments and arcs of gas connecting large-scale, extended gas structures to disk scales. Prevalent late infall has far-reaching consequences for planet formation theory, challenging the long-standing treatment of evolved disks in isolation. In this work, we investigate the emergence of late-infall streamers in different formation scenarios, their morphology and multiplicity, as well as their dependence on environmental conditions. We conducted this investigation by performing 3D hydrodynamical simulation using the grid-based code \texttt{FARGO3D}, which we post-process to obtain synthetic observations using the Monte Carlo radiative transfer code \texttt{RADMC3D}. We find that, while a late infall event in the form of a single encounter with a "cloudlet" of gas can produce a streamer via an interplay between the fallback of bound material and shocks, such features dissipate quickly, on a timescale of ${\sim}\SI{10}{\kilo\year}$. Furthermore, we find that streamers can also form naturally in a turbulent, dense environment without the need for such encounters, which could act to reconcile short-lived streamers with ubiquitous detection of these structures. Here, we find multiple co-existing streamers for a disk velocity relative to the interstellar medium of $v_\mathrm{sys}=\SI{0.5}{\kilo\meter\per\second}$ and a turbulent velocity dispersion of $\sigma_\mathrm{turb}=\SI{0.5}{\kilo\meter\per\second}$, in which case an angular momentum flux of $\dot{J}\sim\SI{e38}{\gram\centi\meter\squared\per\second\squared}$ arises. We find considerable dependence of the streamer morphology on the environment, which may act as a utility to constrain the physical conditions of the gas surrounding planet-forming disk, and therefore the conditions under which planets form.}


    \maketitle
\section{Introduction}
Planet formation theory has evolved greatly in recent years, with more complete models now connecting the large range of size scales of dust grain growth, the formation of planetary cores, migration, and gas accretion (for a review, see \citealt{drazkowska2023}). Despite the steady advancement, protoplanetary disks, the birthplaces of planets, are still largely treated as being isolated from their environment. After their initial formation, the dust budget is treated as fixed, being converted into planetary cores through internal processes. Issues remain in this framework; in particular, recent observations indicate that the dust mass budget of fully formed protoplanetary disks is insufficient to form the giant planets we observe \citep{manara2018}.

The mass budget problem has sparked an increased interest in treating the possible onset of formation in young, embedded disks still subject to infall \citep{drazkowska2018,morbidelli2022,huehn2025,carrera2025}. However, infall may also play a major role in the treatment of older, fully formed disks \citep{ginski2021,huang2021,garufi2022,pineda2023,gupta2024}, to the point where an entirely new, second-generation disk may be formed \citep{dullemond2019,kuffmeier2020}. The assumption of isolated evolution is especially challenged by the increasing number of disk observations, both in scattered light and molecular line emission, indicating direct interactions with their large-scale environment. A recent census of about half of the observable Class II disks in the Taurus star-forming region with VLT/SPHERE reveals hints of environment interactions for about a third of the sample \citep{garufi2024}. They mostly present in the form of streamers, which are arcs of gas connecting the protoplanetary disk scales with its more diffuse surroundings. Furthermore, some targets show various spiral structures that may be related to these streamers \citep{calcino2025}. Recent studies of such structures in the AB Aur system reveal that, while dynamical modeling of gravitational instability \citep{speedie2024} can explain the observed structure, specific spirals may actually be connected to the larger-scale environment, identifying them as streamers of infalling material. The detection of SO line emission, believed to be a shock tracer, further supports this hypothesis \citep{speedie2025}.

In the past, studies have been conducted to understand the nature of environment-disk interactions. In particular, the structures around the HL Tau and DG Tau disks have been confirmed as inflowing material associated with an accretion shock \citep{garufi2022}, and the dynamics of the flow onto DG Tau has been explained by models of cloudlet accretion \citep{hanawa2024}. Similarly, late infall has been confirmed in SU Aur \citep{ginski2021}, which also shows a self-shadowing structure that may be related to a misaligned inner disk; a peculiar configuration that may be caused by the observed infall. Recently, there have been an increasing number of detections of misaligned and warped inner disks. \citet{villenave2024} showed that 75\% of the targeted edge-on disks show asymmetries that are likely caused by a tilted inner region. Because streamers have been observed in such systems, infall of material with misaligned angular momentum is a prime candidate to explain their origin \citep{kuffmeier2021,dullemond2022}. In turn, this would indicate that the continuous infall of material, even at later evolutionary stages, is a ubiquitous phenomenon.

This hardening observational evidence of evolved disks interacting with and accreting mass from their environment raises the important question of the origin and nature of these structures. Streamers may arise as a connection from clearly defined mass reservoirs at larger scales down to disk scales \citep{kuffmeier2023}, modeled on smaller scales in the form of "cloudlets" \citep{dullemond2019,kuffmeier2021}, or from interactions in a system with a stellar companion or a flyby \citep{dai2015,kurtovic2018,pfalzner2024}. However, they are detected around a wide range of disk-hosting stars, and the possible rarity of encounters with cloudlets of considerable density or other stars might disfavor such interactions as the underlying reason. Alternatively, they may arise solely as the result of interactions with the diffuse interstellar medium (ISM). Here, material is continuously supplied via Bondi-Hoyle-Lyttleton (BHL) accretion \citep{krumholz2006,padoan2025,pelkonen2025}, steadily supplying mass and angular momentum that naturally creates streaming structures without the need for a distinct source. This simple process can reproduce various observed disk properties \citep{winter2024}, suggesting that it may offer an explanation for the high abundance of observed inflow structures.

Understanding the mechanisms that can create streamer-like structures and shape their morphology, as well as identifying their potential origin, is vital for our understanding of the dynamical and mass-budget evolution of protoplanetary disks, with potential far-reaching consequences for planet formation as a whole. In this work, we investigate two different types of interactions of fully formed protoplanetary disks with their environment using hydrodynamical simulations. First, we probed how a single, well-characterized interaction with a cloudlet of gas can create various streamer-like observational signatures (Section \ref{sec:res_cloudlet}). We then considered a disk that is embedded in the turbulent interstellar medium to examine structures that arise in the absence of a distinct source of infall (Section \ref{sec:res_bhl}). While signatures of single interactions are not long-lived, those arising from constant interactions with the ISM could persist over longer time scales and would therefore be much more abundantly observable.

\section{Methods}\label{sec:methods}
We performed 3D hydrodynamical simulations on a spherical grid with logarithmic spacing in the radial direction, and uniform spacing in the polar and azimuthal direction. The computations were performed using the \texttt{FARGO3D} code \citep{benitez-llambay2016}, capable of GPU multiprocessing while utilizing the advection algorithm by \citet{masset2000}. The radial extent of the grid was from $r_\mathrm{min}=\SI{10}{\astronomicalunit}$ to $r_\mathrm{max}=\SI{5000}{\astronomicalunit}$, except the simulation presented in Section \ref{sec:res_cloudlet}, where $r_\mathrm{min}=\SI{5}{\astronomicalunit}$. We moved the outer edge further outward to $r_\mathrm{max}=\SI{50000}{\astronomicalunit}$ for simulations with a turbulent initial condition, described in later sections. In the polar direction, the range was from $\theta_\mathrm{min}=\SI{10}{\degree}$ to $\theta_\mathrm{max}=\SI{170}{\degree}$. In the radial and polar direction, we employed an outflow boundary condition, whereas the azimuthal boundary is periodic. There are $N_r=250$ cells in the radial, $N_\theta=100$ cells in the polar and $N_\phi=225$ cells in the azimuthal direction, resulting in $H/\Delta r = H/(r\Delta\theta)=H/(r\Delta\phi)\approx 1.3$ at $r=\SI{5.2}{\astronomicalunit}$ (${\approx}3$ at \SI{100}{\astronomicalunit}). Here, $H$ is the gas pressure scale height and $\Delta r$, $\Delta\theta$, and $\Delta\phi$ are the resolution in the respective direction. We utilized an isothermal equation of state, $P=c_s^2\rho$, where $P$ is the gas pressure, $c_s$ is the isothermal sound speed, and $\rho$ is the gas density. A $\SI{1}{\solarmass}$ star was placed at the center of the grid. Additionally, we added artificial viscosity to the simulation, corresponding to an $\alpha$ parameter \citep{ss1973} of $\alpha=\num{e-3}$. Due to the lack of self-gravity in our simulation, we deliberately omitted indirect terms rooted in the stellar acceleration caused by the gas in the potential.

\subsection{Protoplanetary disk setup}
We introduced a protoplanetary disk around the star at the center of our simulation, so that the midplane coincides with the equatorial plane. We set the density of the corresponding grid cells to
\begin{equation}
    \begin{split}
    \rho(r,\theta) &= \frac{\Sigma_0}{\sqrt{2\pi}h_0R_0}\left(\frac{r_\mathrm{cyl}(r, \theta)}{R_0}\right)^{-a-b-1}\\
    &\times\exp\left(-\frac{z(r, \theta)^2}{2h_0^2R_0^2}\left(\frac{r_\mathrm{cyl}(r,\theta)}{R_0}\right)^{-2b-2}\right)\\
    &\times\left(1+\exp\left(\frac{r_\mathrm{cyl}(r,\theta)-R_c}{0.05R_c}\right)\right)^{-1},
    \end{split}
\end{equation}
where $a$ is the surface density slope, $b$ is the flaring index, $r_\mathrm{cyl}$ and $z$ are the cylindrical coordinates associated with position $(r,\theta)$ in the spherical coordinate system, $R_c$ is the disk cutoff radius, $R_0=\SI{5.2}{\astronomicalunit}$ is a reference radius, $h_0$ is the aspect ratio at that reference radius, and $\Sigma_0$ is the surface density at that radius. For the disk itself, we chose $\Sigma_0=\SI{310.547}{\gram\per\centi\meter\squared}$, $a=1.5$, and $R_c=\SI{100}{\astronomicalunit}$, resulting in a disk with a mass of $M_d=\SI{0.05}{\solarmass}$. In the interest of numerical stability, we introduced a floor value for the density, $\rho_\mathrm{floor}=\SI{e-22}{\gram\per\centi\meter\squared}$.

We defined the temperature such that the aspect ratio $h=H/r$, where $H$ is the pressure scale height, is described by
\begin{equation}
    h(r,\theta) = h_0\left(\frac{r_\mathrm{cyl}(r,\theta)}{R_0}\right)^b.
\end{equation}
To represent the case of passive irradiation by a star with solar luminosity, we used $h_0=0.03799$ and $b=0.25$. Furthermore, we included a floor temperature of $T_\mathrm{floor}=\SI{10}{\kelvin}$, resulting in the isothermal sound speed being given by
\begin{equation}
    c_s(r,\theta) = \left(\left(h(r, \theta)r_\mathrm{cyl}(r,\theta)\Omega\right)^8+\left(\frac{k_BT_\mathrm{floor}}{\mu}\right)^4\right)^{1/8},
\end{equation}
where $k_B$ is the Boltzmann constant, $\Omega$ is the Keplerian orbital frequency, and $\mu=2.3m_p$ is the mean molecular weight of the gas, with $m_p$ the proton mass.

To take the effects of the gas pressure gradient and the exponential cutoff of the disk into account, we initialized our simulations with an azimuthal velocity that deviates from the Keplerian value,
\begin{equation}
    \begin{split}
    v_\phi(r,\theta)^2 &= \Omega^2r_\mathrm{cyl}^2+c_s^2\\
    &\times\left((b+1)\left(\frac{z(r,\theta)}{r_\mathrm{cyl}(r,\theta)h}\right)^2\right.\\
    &\left.-\frac{r_\mathrm{cyl}(r,\theta)}{0.05R_c\left(1+\exp\left(-\frac{r_\mathrm{cyl}(r,\theta)-R_c}{0.05R_c}\right)\right)}-a+b-2\right),
    \end{split}
\end{equation}
which can become imaginary, in which case we set $v_\phi=\Omega r_\mathrm{cyl}$. We set the other two velocity components to zero, $v_r=v_\theta=0$.

\subsection{Setup of the environment}
We simulated the late accretion of gas onto Class II disks in two different scenarios. First, we considered a controlled case where the environmental interactions are represented by a single, spherical cloudlet of gas encountering the disk on a hyperbolic orbit. Subsequently, we considered the more realistic case of a disk with a systemic velocity embedded in a turbulent interstellar medium (ISM), utilizing a compressively turbulent initial condition with a given power spectrum and dispersion.

\subsubsection{Spherical cloudlet}\label{sec:methods_cloudlet}
The center of the initially spherical cloudlet was placed at a position described by the initial distance $d_0$ and the impact parameter $b$, as well as the speed at infinity $v_\infty$ that sets the critical impact parameter $b_\mathrm{crit}=\sqrt{GM_\star/v_\infty^2}$. The initial position ($x_0,y_0$) is then described by two values in Cartesian coordinates,
\begin{align}
    \nu &= \arccos\left(\frac{\frac{b^2}{b_\mathrm{crit}}-d_0}{d_0\sqrt{1+\left(\frac{b}{b_\mathrm{crit}}\right)^2}}\right),\\
    x_0 &= d_0\cos(\nu),\\
    y_0 &= d_0\sin(\nu),
\end{align}
with $\nu$ the true anomaly. The initial velocity is given by
\begin{align}
    v_{x,0} &= v_\infty\sqrt{\frac{2b_\mathrm{crit}}{d_0}+1}\nonumber\\
    &\times\cos\left(\arctan\left(\frac{\sin(\nu)}{\cos(\nu)+\left(1+\left(\frac{b}{b_\mathrm{crit}}\right)^2\right)^{-1/2}}\right)-\nu+\frac{\pi}{2}\right),\\
    v_{y,0} &= v_\infty\sqrt{\frac{2b_\mathrm{crit}}{d_0}+1}\nonumber\\
    &\times\sin\left(\arctan\left(\frac{\sin(\nu)}{\cos(\nu)+\left(1+\left(\frac{b}{b_\mathrm{crit}}\right)^2\right)^{-1/2}}\right)-\nu+\frac{\pi}{2}\right),
\end{align}
which was applied to all grid cells that belong to the cloudlet, thereby neglecting the spatial extent of it for the calculation of the orbital velocity. We finally rotated the cloudlet's position and velocity vectors by 30 degrees around a hypothetical $x$-axis to achieve an out-of-plane infall initial condition.

The cloudlet was initialized with $v_\infty=\SI{0.5}{\kilo\meter\per\second}$, $d_0=\SI{3000}{\astronomicalunit}$, a radius of $R_\mathrm{cloud}=\SI{500}{\astronomicalunit}$, and a mass of $M_\mathrm{cloud}=\SI{5e-3}{\solarmass}=0.1M_d$, resulting in an initial density of $\rho_\mathrm{cloud,0}=\SI{5.67e-18}{\gram\per\centi\meter\cubed}$. We chose $b=0.5b_\mathrm{crit}$ for the impact parameter. The mass ratio was chosen such that the disk disruption due to the cloudlet is moderate. We selected the other parameter values with the goal to keep the radial domain of the grid and the physical time until the interaction small, and ensure $b<b_\mathrm{crit}$ across the entire physical extent of the cloudlet. The mass of the cloudlet is higher than suggested by observational data from giant molecular clouds \citep{falgarone2004,kh2010} for our choice of initial radius, but the lack of pressure support from a warm ISM leads to expansion of the cloudlet before encountering the disk. This initial condition was not intended to directly represent realistic physical conditions. Rather, we started with a more compact cloudlet to keep the inaccuracy incurred by neglecting the physical extent when initializing the orbital velocity small. When initially encountering the disk, the mean density of the cloudlet gas is ${\sim}\SI{e-19}{\gram\per\centi\meter\cubed}$, which is comparable to the density of filamentary structures in the Taurus star forming region \citep{pineda2010,palmeirim2013}. This initial expansion does not affect the angular momentum of the cloudlet relative to the disk. Therefore, we do not expect it to substantially affect our results. The total runtime of the cloudlet simulation is $\SI{42.7}{\kilo\year}$, after which any arising structures have dissipated.

\subsubsection{Turbulent interstellar medium}\label{sec:methods_bhl}
The turbulent ISM simulations were run without a cloudlet, but with a global density $\rho_\mathrm{ISM}$ and turbulent initial velocity field, modeled as a Gaussian random field \citep{dubinski1995,krumholz2006,seifried2013}. The density was determined via the systemic infall rate, $\dot M_\mathrm{sys}$, which is a free parameter, resulting in
\begin{equation}
    \rho_\mathrm{ISM} = \frac{\dot M_\mathrm{sys}}{\pi R_\mathrm{bondi}^2v_\mathrm{sys}},
\end{equation}
where $R_\mathrm{bondi}=2GM_\star/c_s^2$ is the Bondi accretion radius of the star, $M_\star$ is the stellar mass, $c_s$ is the isothermal sound speed of the ISM, and $v_\mathrm{sys}$ is the absolute value of the systemic velocity of the star-disk system as it moves through the ISM. Unless specified otherwise, we adopted $\dot M_\mathrm{sys}=\SI{e-8}{\solarmass\per\year}$, so that $\rho_\mathrm{ISM}=\SI{1.5e-21}{\gram\per\centi\meter\cubed}$ for $v_\mathrm{sys}=\SI{1}{\kilo\meter\per\second}$.

The velocity field was realized by randomly sampling $N_\mathrm{wav}=10000$ sets of the wave number $k_n=\left|\vec k_n\right|$, the wave propagation unit vector $\hat{k_n}$, the components of the corresponding velocity $\vec v_k$, and a phase $\phi_n$. We sampled $k$ logarithmically in the interval $[k_\mathrm{min}, k_\mathrm{max})$, $\hat{k}$ on a spherical surface with uniformly chosen $\phi_k\in[0,2\pi)$ and $\mu_k=\cos\theta_k\in[-1,1)$, and $\phi_n\in[0,2\pi)$. We chose a Gaussian random field with a power spectrum $P(k)\sim k^{-4}$ as the initial condition, achieved by sampling the components of $\vec v_k$ from a Gaussian distribution with
\begin{equation}
    \sigma_k = \sqrt{\frac{k_\mathrm{min}k_\mathrm{max}}{N_\mathrm{wav}(k_\mathrm{max}-k_\mathrm{min})}\log\left(\frac{k_\mathrm{max}}{k_\mathrm{min}}\right)}\sigma_\mathrm{turb}k^{-0.5},\label{eq:sigma_v}
\end{equation}
where $\sigma_\mathrm{turb}$ is the standard deviation of the total velocity field in real space, which is a model parameter. The turbulent velocity field at location $\vec r$ is then given by
\begin{equation}
    \vec v_\mathrm{turb}(\vec r) = \sum_{n=0}^{N_\mathrm{wav}}\vec v_k\cos(k_n\hat{\vec k}_n\cdot\vec r + \phi_n).
\end{equation}

We added systemic infall, caused in reality by a systemic velocity, $\vec v_\mathrm{sys}$, to mimic the movement of the system through the ISM, so that the total gas velocity is given by
\begin{equation}
    \vec v(\vec r) = \vec v_\mathrm{turb}(\vec r) + v_\mathrm{sys}\hat{\vec v}_\mathrm{sys}(\vec r),
\end{equation}
where we chose $\hat{\vec v}_\mathrm{sys}=(0,\frac{1}{\sqrt{2}},\frac{1}{\sqrt{2}})$ as the unit vector for the direction of systemic infall to consider one realization of the expected random alignment between the disk and systemic velocity. The velocity initial condition is therefore described by the two parameters $\sigma_\mathrm{turb}$ and $v_\mathrm{sys}$, with the two parameters $k_\mathrm{min}$ and $k_\mathrm{max}$ describing the limits of the turbulence power spectrum. We note that the turbulence was only introduced as an initial condition, and we did not apply turbulent forcing to drive the turbulence dynamically. As a result, the simulation is no longer self-consistent after the gas from the outer grid edge has reached the disk. At the highest systemic velocity we considered, $v_\mathrm{sys}=\SI{1}{\kilo\meter\per\second}$, this occurs after $\SI{237}{\kilo\year}$, which is much larger than the employed simulation runtime of \SI{30}{\kilo\year}. Nevertheless, the lack of turbulent forcing results in a decrease of the turbulent velocity dispersion over time. In terms of the turbulence Mach number $\mathcal{M}$, a turbulent velocity diffusion of $\sigma_k=\{1,0.5,0.25,0.1\}\ \si{\kilo\meter\per\second}$ corresponds to $\mathcal{M}=\{8,4,3.2,2.9\}$ at the beginning of the simulation, which has decayed down to $\mathcal{M}=\{6,3,2.9,2.8\}$ in the end.

\subsection{Synthetic observations}
Using the Monte Carlo radiative transfer code \texttt{RADMC3D} \citep{radmc3d}, we created synthetic observations of select simulation snapshots. We produced observations of the CO J 2--1 transition line, assuming a constant CO to $H_2$ mass ratio of \num{e-4}, using 250 points in frequency space covering a spectral range from $-\SI{5}{\kilo\meter\per\second}$ to $+\SI{5}{\kilo\meter\per\second}$ to create maps of the first moment and the peak brightness temperature, $T_\mathrm{b,peak}$. To calculate $T_\mathrm{b,peak}$, we used the Rayleigh-Jeans approximation. We did not employ any prescriptions for the freeze-out or photo-dissociation of CO. Instead of the isothermal temperature distribution employed in the simulations, we first re-computed the temperature with a separate Monte Carlo simulation with $N_\mathrm{phot}=\num{e8}$ photons, where we considered the central star to have a blackbody spectrum of a Solar mass star with Solar luminosity. The wavelength grid consists of 100 wavelengths between $\lambda_\mathrm{min}=\SI{0.09}{\micro\meter}$ and $\lambda_\mathrm{max}=\SI{3000}{\micro\meter}$.

We also created synthetic scattered-light observations. Here, the required dust opacities were calculated using the \texttt{optool} code \citep{dominik2021}, where we assumed the DIANA standard dust model composition \citep{woitke2016}. The grains are composed of 70\% Mg pyroxene and carbon, which are mixed with a mass ratio of 87\% and 13\%, respectively. We assumed a porosity of 25\% and mix 15 different grain sizes between $\SI{1}{\micro\meter}$ and $\SI{3}{\micro\meter}$, sampling an $n(a)\ \propto\ a^{2.5}$ power-law size distribution. Despite only including such small grains, we assumed a dust-to-gas volume density ratio of \num{e-2}, assuming that grains in the infall structures cannot grow until reaching the disk. This assumption breaks down inside the disk itself, which we neglect as we focus on the inflow structures themselves. We considered full polarization-dependent scattering, though we flattened the peak of the scattering phase function to avoid numerical issues by ``chopping'' at \SI{2}{\degree}. Using this technique, photons scattered at an angle below this value relative to the forward scattering direction are instead treated as not having interacted at all \citep{dominik2021}. For the spatial distribution of the dust, we assumed that it perfectly follows that of the gas, with a dust-to-gas ratio of 1\%.

The scattering Monte Carlo simulation was performed by tracing $N_\mathrm{phot,scat}=\num{e9}$ photons. The image was taken at a wavelength of $\lambda_\mathrm{img}=\SI{1.245}{\micro\meter}$, which we used to calculate the absolute value of the $Q_\phi$ polarization,
\begin{equation}
    Q_\phi = \left|Q\cos(2\phi)+U\sin(2\phi)\right|,
\end{equation}
where $\phi$ is the azimuthal position on the image plane and $Q$ and $U$ are the polarization Stokes vector components.

\section{Results}
\begin{figure*}[htp]
    \centering\includegraphics[width=.8\linewidth]{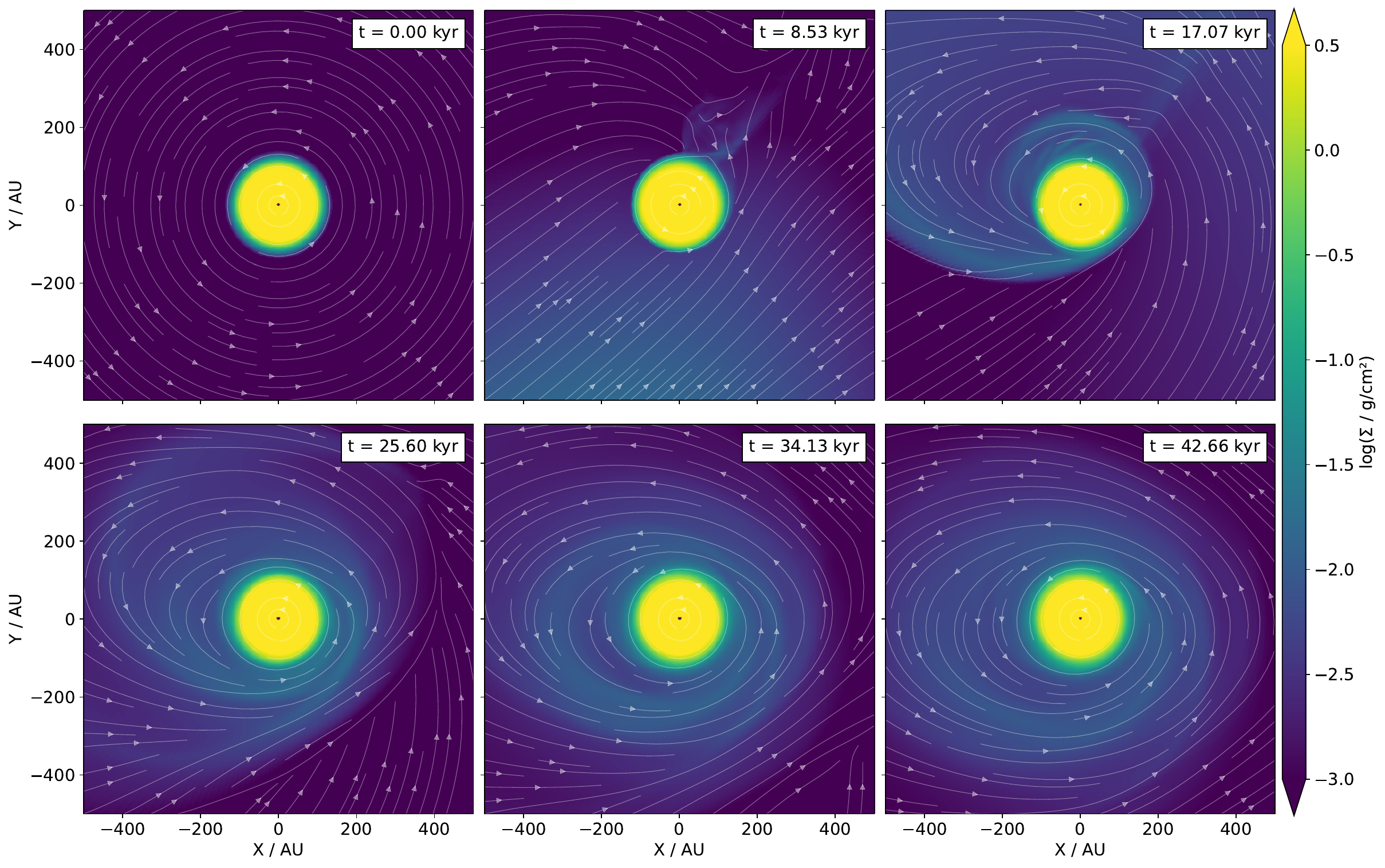}
    \caption{Gas column density of the cloudlet encounter simulation, for six different points in time. The perspective is face-on. The white arrows show the velocity streamlines, computed as a mass-weighted average along the line of sight. The color scale was chosen to highlight low-density features, thereby saturating the protoplanetary disk.}
    \label{fig:si_sigma_t}
\end{figure*}%
In this work, we investigated the emergence of streamer-like structures in two different scenarios. First, we considered the case of a single gas cloudlet being accreted onto the disk (see Sect. \ref{sec:methods_cloudlet}). The primary advantage of this scenario is that it is a controlled environment with a well-defined encounter, however, we lack realism. Second, we considered the more realistic case of a star-disk system moving through the dense, turbulent ISM (see Sect. \ref{sec:methods_bhl}), which inherently contains more randomness.

\subsection{Cloudlet capture}\label{sec:res_cloudlet}
In the cloudlet capture scenario, we find that the arising structures occur for a fundamentally different reason than in previously considered model setups (e.g., \citealt{dullemond2019,hanawa2024}). With passive stellar irradiation being the only heating source, the background ISM surrounding the initially spherical cloudlet is cold, so that it lacks pressure support and expands freely. If, on the contrary, the ISM was modeled as warm as in previous models, it would act to confine the cloudlet. Differently from our scenario, the arced structure of a streamer would be the natural outcome of a spherical cloud being deformed as the gas elements move along their orbit.

As a result of this difference, the cloudlet has already expanded considerably at the time when it encounters the disk. This is apparent in the second panel of Fig. \ref{fig:si_sigma_t}, showing the time evolution of the surface density in and around the disk, as well as the flow lines of the gas. Indeed, in the given field of view, the original spherical geometry is fully lost, and it can be seen that diffuse gas engulfs the disk fully during the encounter. As a direct consequence, over-shooting mass is accumulated on the side of the disk opposite to the infall direction. This morphology is reminiscent of an accretion tail that is the result of converging flow lines caused by gravitational focussing. A key difference to such an accretion tail lies in the angular momentum the accumulated material has with respect to the star. The hyperbolic orbit of the cloudlet has an impact parameter of $b=0.5b_\mathrm{crit}$, and angular momentum is not lost in the expansion process; instead, it is reflected by the asymmetry of the density and velocity at the time of the encounter.

The gas that is accumulated in the accretion tail remains bound to the star, because $b<b_\mathrm{crit}$. Due to the angular momentum it has with respect to the disk, it falls back onto it in an arced shape. This "fallback streamer" is supported in its shape by the flow of material that is still inflowing from the original direction, creating a shock at the edge of the arc. The resulting streamer morphology can be seen in the third panel of Fig. \ref{fig:si_sigma_t}. Due to being caused by the fallback of material that was accumulated in the accretion tail, we find that the direction of the streamer differs from the original inflow direction.

However, the resulting streamer is short-lived. Only ${\sim}\SI{8.5}{\kilo\year}$ later, the surface density of the fallback streamer has decreased in contrast considerably, as shown in the forth panel of Fig. \ref{fig:si_sigma_t}. Subsequently, the distribution of mass in the simulation can no longer be recognized as a streamer, although left-over material serves as evidence of the encounter until it is eventually accreted on timescales beyond the runtime of the simulation.

\begin{figure}[htp]
    \centering\includegraphics[width=.9\linewidth]{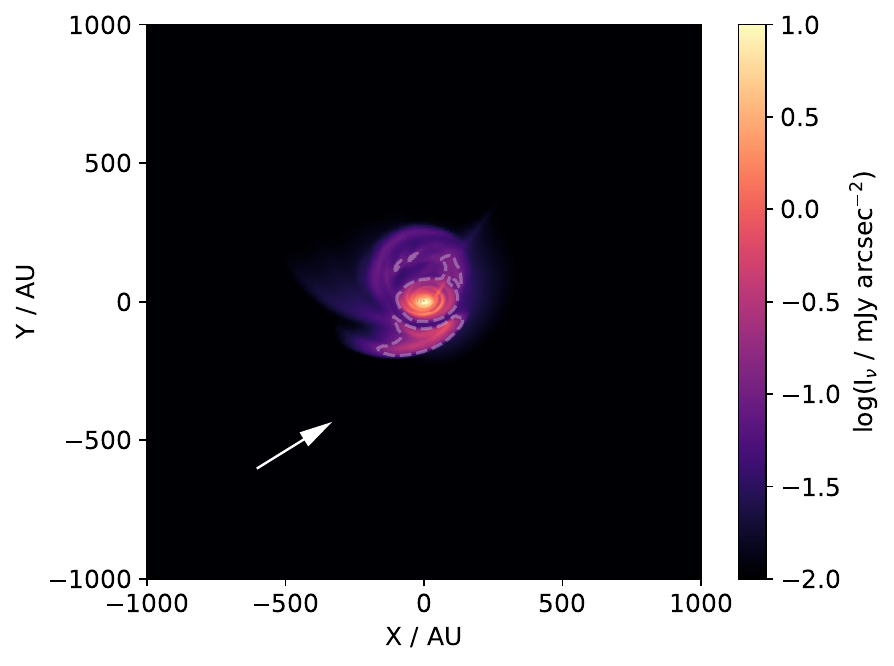}
    \caption{Polarized scattered light ($\lambda=\SI{1.245}{\micro\meter}$) intensity at $t=\SI{17.07}{\kilo\year}$ for the cloudlet encounter simulation. The inclination of the camera is $i=\SI{40}{\degree}$. The white arrow represents the original orbital velocity of the cloudlet, and the white dashed line is the \SI{0.1}{\milli\jansky\per\arcsecond\squared} contour line.}
    \label{fig:si_scattered}
\end{figure}%
We investigated the fallback shock streamer using a synthetic polarized scattered-light observation, shown in Fig. \ref{fig:si_scattered}. We find that it produces a bright, extended, and arced signal at a wavelength of $\lambda=\SI{1.245}{\micro\meter}$, so that it would indeed be recognized as a streamer in observations, though only the bottom part close to the disk has a flux ${>}\SI{0.1}{\milli\jansky\per\arcsecond\squared}$. Additionally, the inflowing material causes spiral structures to arise at the disk surface, visible in the scattered light. Such a combination of a streamer and spiral structures is found in systems though to be undergoing infall, like SU Aur \citep{ginski2021}.

\begin{figure}[htp]
    \centering\includegraphics[width=\linewidth]{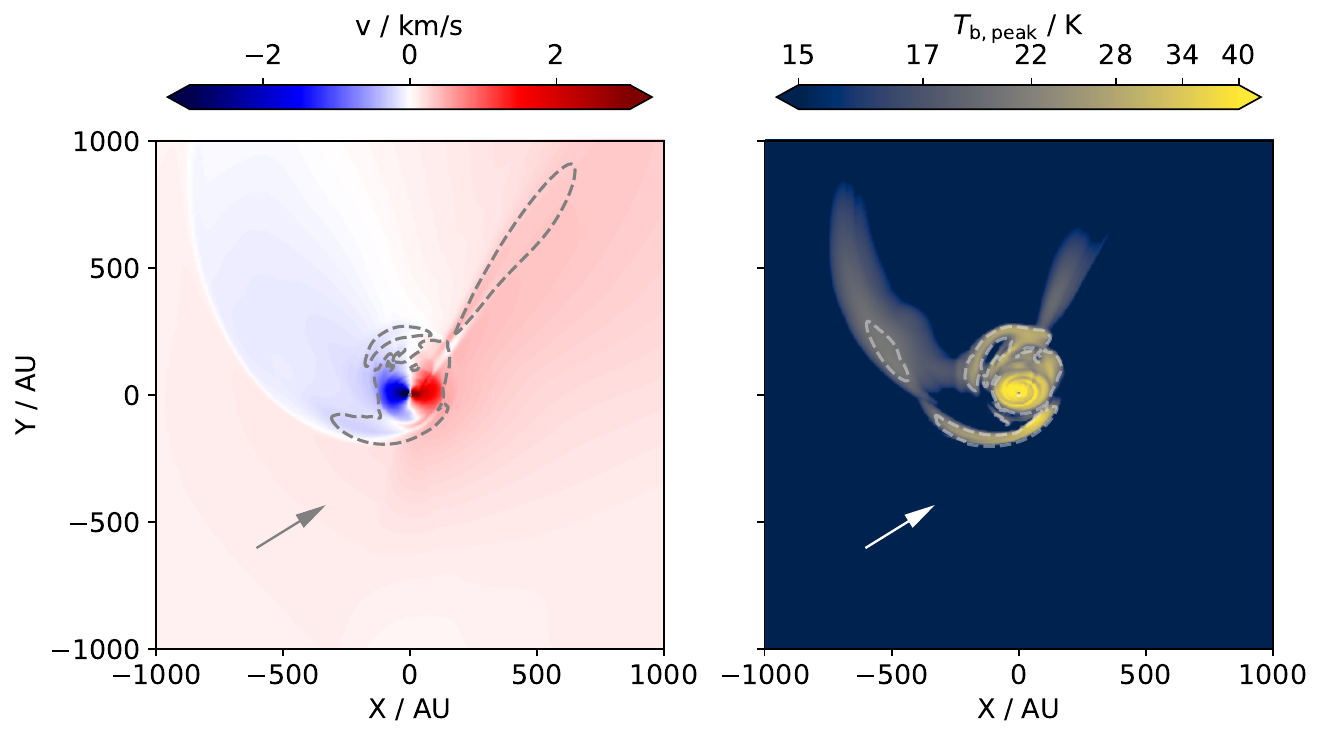}
    \caption{Synthetic CO line emission (J 2-1) at $t=\SI{17.07}{\kilo\year}$ for the cloudlet encounter simulation. The left panel shows the moment 1 map, with the gray dashed line showing the $\SI{600}{\milli\jansky\per\arcsecond\squared\kilo\meter\per\second}$ moment 0 contour. The right panel shows the peak brightness temperature on a square root scale. The gray and white arrows represent the initial orbital velocity of the cloudlet. The white dashed line in the right panel represents the $T_\mathrm{b,peak}=\SI{20}{\kelvin}$ contour line.}
    \label{fig:si_co}
\end{figure}%
Furthermore, we created synthetic observations of the CO J 2-1 transition. The resulting moment 1 and peak brightness temperature maps are shown in Fig. \ref{fig:si_co}. The streamer can be recognized dynamically, adding a recognizable non-Keplerian component to the moment 1 map, similar to systems with observed streamers like HL Tau and DG Tau \citep{garufi2022}. Closer to the disk, the moment 0 signal of the shock fallback streamer exceeds $\SI{600}{\milli\jansky\per\arcsecond\squared\kilo\meter\per\second}$, and the arc visible in scattered light is also clearly visible in the peak brightness temperature map. In the case of DG Tau, the channel map morphology was explained using a model of a cloudlet confined by a warm medium \citep{hanawa2024}; here, we find that the accretion of a more extended gas structure, if it has angular momentum relative to the star, can lead to similar results.

\begin{table}[htp]
    \centering
    \caption{Simulation parameters for accretion in turbulent medium}
    \begin{tabular}{lccccc}
        \hline
        \hline
        No. & $v_\mathrm{sys}$ (\si[per-mode=symbol]{\kilo\meter\per\second}) & $\sigma_\mathrm{turb}$ (\si[per-mode=symbol]{\kilo\meter\per\second}) & $k_\mathrm{max}$ (\si[per-mode=symbol]{\per\astronomicalunit}) & $\dot M_\mathrm{sys}$ (\si[per-mode=symbol]{\solarmass\per\year})\\
        \hline
        1 & 1 & 1 &  $2\pi/50$ & \num{e-8}\\
        2 & 0.5 & 0.5 & $2\pi/50$ & \num{e-8}\\
        3 & 0.5 & 0.5 & $2\pi/1000$ & \num{e-8}\\
        4 & 0.5 & 0.25 & $2\pi/50$ & \num{e-8}\\
        5 & 0.5 & 0.1 & $2\pi/50$ & \num{e-8}\\
        6 & 0.5 & 0.5 & $2\pi/50$ & \num{e-9}\\
        \hline
    \end{tabular}
    \label{tab:bh_params}
\end{table}%
\subsection{Accretion in a turbulent medium}\label{sec:res_bhl}
The emergence of fallback shock streamers in the previous section is the result of the encounter with an initially spherical gas cloudlet. Even though it has evolved to a more extended structure by the time it reaches the star, the initial condition is representative of a considerable over-density in the ISM with a symmetric shape. However, it remains an open question how frequent such encounters with clearly discernible cloudlets of gas are, and how well they describe the conditions of the large-scale environment of a star-forming region. Even if they were to serve as a reasonable approximation, we find that the resulting streamer is short-lived. If encounters of this type were the dominant cause for streamers, they should be much rarer than the observational sample suggests \citep{garufi2024,villenave2024}.

In this section, we offer a way to reconcile this discrepancy. In an effort to model the physical conditions of a star-forming region more realistically, we instead modeled the star-disk system as moving through the turbulent interstellar medium with a given systemic velocity (see Sect. \ref{sec:methods_bhl}). The resulting total velocity field is shown in Appendix \ref{app:turbfield}

In this scenario, streamers can form in a way qualitatively similar to the cloudlet capture case. Rather than creating a reservoir of gas with angular momentum relative to the disk by overshooting cloudlet material, it arises naturally in the form of a BHL accretion tail \citep{hl1939,bondi1945}. In the absence of turbulence, this accretion tail cannot lead to the formation of arc-shaped structures, as no asymmetries that would create angular momentum with respect to the disk would be present. As a result, the material would fall back on a straight trajectory. In the presence of turbulence, however, such asymmetries can naturally be created randomly. Even if the flow structure was unperturbed by the velocity, angular momentum could result from spontaneous enhancements in density. In a simplified picture, one might imagine the large-scale flow lines to vary in direction between two points in time, thereby creating, on average, a qualitatively similar picture as the extended structure caused by the encounter with an expanded cloudlet.

While this idea can help to understand the possibility of fallback streamers arising, the full turbulent picture is more complex. In the following, we discuss simulation results from multiple simulations with various resulting structures. We investigated how the arising structures change with $v_\mathrm{sys}$, $\sigma_\mathrm{turb}$, and $k_\mathrm{max}$. We considered the influence of $\dot M_\mathrm{sys}$ in Appendix \ref{app:low_acc}. The numerical choices for these parameters are shown in Table \ref{tab:bh_params}.

\begin{figure}[htp]
    \centering\includegraphics[width=0.9\linewidth]{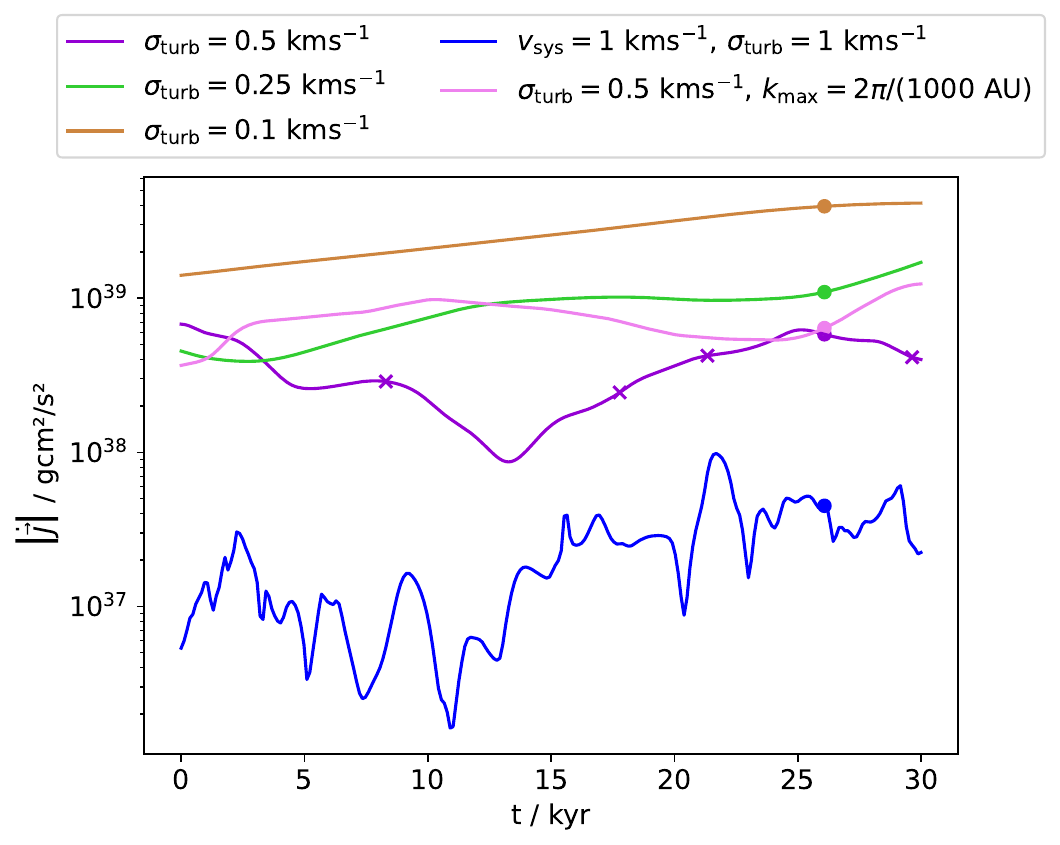}
    \caption{Absolute value of the angular momentum flux through a spherical shell with radius $R_\mathrm{bound}$ as a function of time. The differently colored lines denote different simulations from Table \ref{tab:bh_params}. Unless stated otherwise, the shown simulations have $v_\mathrm{sys}=\SI{0.5}{\kilo\meter\per\second}$ and $k_\mathrm{min}=2\pi/\SI{50}{\astronomicalunit}$. The round markers show the physical time of the snapshot used for the synthetic observations. The cross markers show the physical time of the snapshots used for the time series in Fig. \ref{fig:b_0.5_0.5_d2_t}.}
    \label{fig:b_am}
\end{figure}%
\begin{figure*}[htp]
    \centering\includegraphics[width=0.8\linewidth]{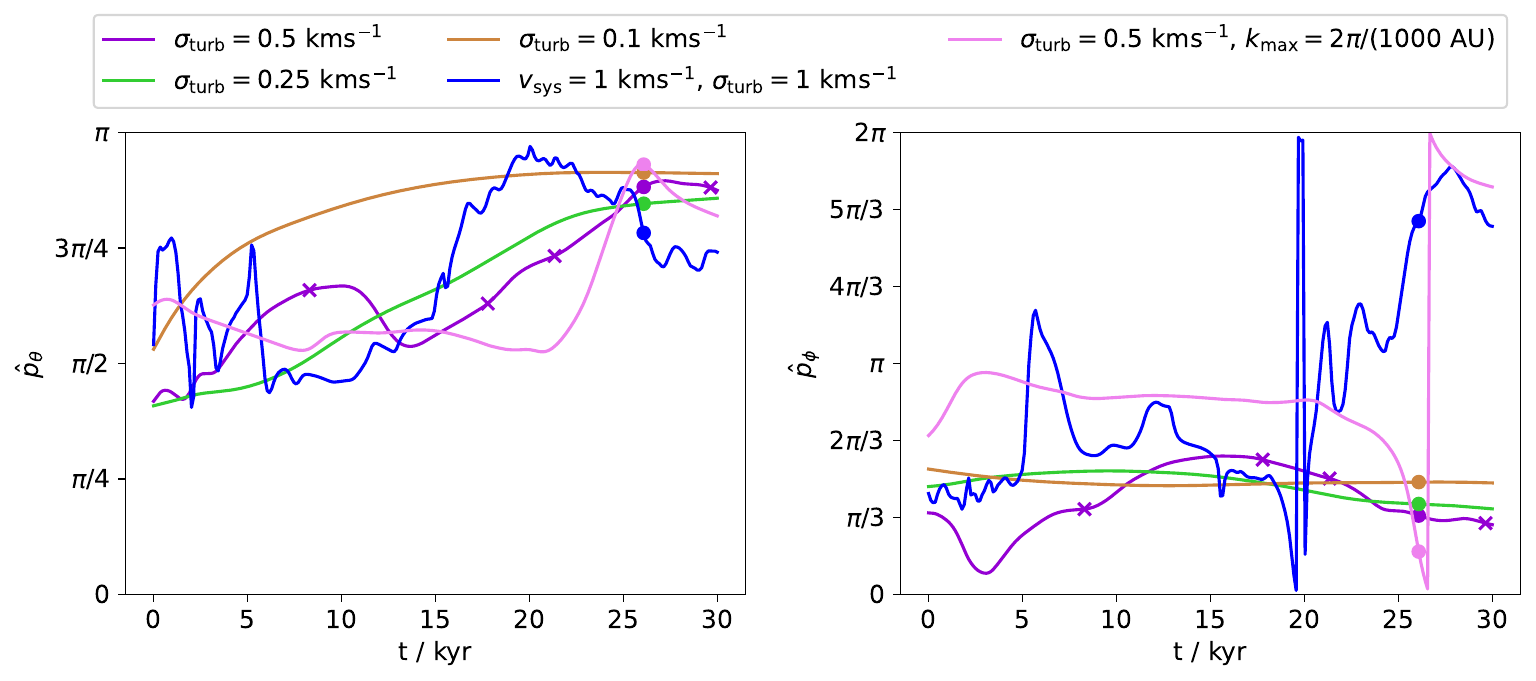}
    \caption{Orientation of the non-systemic momentum flux through a spherical shell with radius $R_\mathrm{bound}$ as a function of time. The differently colored lines denote different simulations from Table \ref{tab:bh_params}. Unless stated otherwise, the shown simulations have $v_\mathrm{sys}=\SI{0.5}{\kilo\meter\per\second}$ and $k_\mathrm{min}=2\pi/\SI{50}{\astronomicalunit}$. The round markers show the physical time of the snapshot used for the synthetic observations. The cross markers show the physical time of the snapshots used for the time series in Fig. \ref{fig:b_0.5_0.5_d2_t}. The left panel shows the azimuthal angle $\theta$, and the right panel shows the polar angle $\phi$.}
    \label{fig:b_mom}
\end{figure*}%
In our analysis, we find that the morphology of the emerging infall structures is predominantly dependent on the flux of material into the region where the ISM gas has low-enough velocity to be initially bound to the star, and the resulting interaction. Gas originating from farther out regions is too energetic to remain bound to the disk, so that no fallback occurs and no fallback streamers can emerge. Neglecting the spatial details of the turbulent velocity field, we approximated this region as a sphere with radius $R_\mathrm{bound}$, given by
\begin{equation}
    R_\mathrm{bound} = \frac{2GM_\star}{(v_\mathrm{sys}+\sigma_\mathrm{turb})^2}.
\end{equation}
For all the discussed simulations, we show the absolute value of the angular momentum flux through the surface of this sphere in Fig. \ref{fig:b_am}. Additionally, the streamer morphology is influenced by the direction of the non-systemic, that is, turbulent gas inflow from the ISM. We describe this quantity by the direction of the linear momentum flux through the spherical surface, shown in Fig. \ref{fig:b_mom}. The systemic and other symmetrically out- and inflowing contributions to this linear momentum flux are zero. The resulting quantities, although limited to a spatially averaged view, are used in following sections in the discussion of the arising streamers. In the interest of brevity, we only discuss one snapshot for most simulations, at $t=\SI{26.07}{\kilo\year}$.

\subsubsection{High systemic velocity: Weakly bound gas accretion}
\begin{figure*}[htp]
   \centering\includegraphics[width=.9\linewidth]{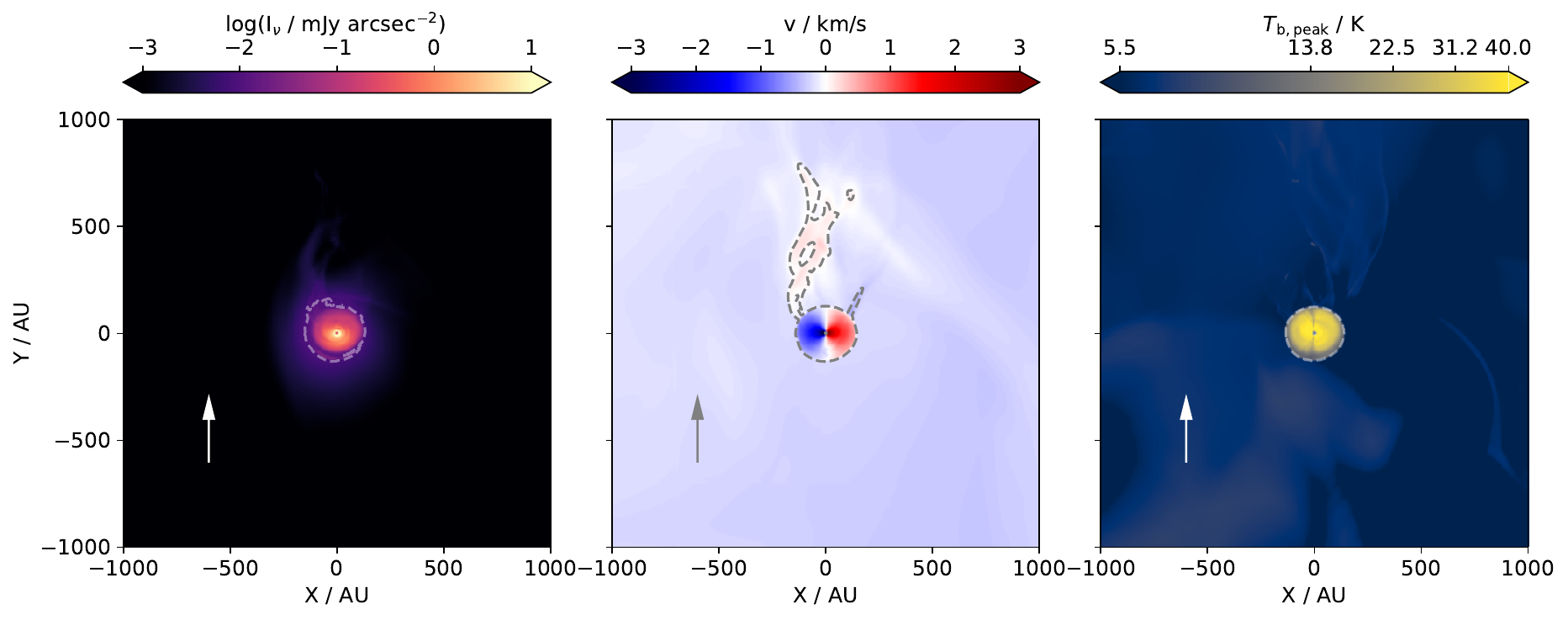}
   \caption{Synthetic observations of simulation number 1 in Table \ref{tab:bh_params}. The left panel shows the polarized scattered light intensity, as in Fig. \ref{fig:si_scattered}, but at an inclination of $i=\SI{30}{\degree}$. The middle and right panels show the CO emission moment 1 and peak brightness temperature maps, as in Fig. \ref{fig:si_co}, but with $i=\SI{30}{\degree}$. The peak brightness temperature values are mapped to colors on a square root scale. Here, the gray dashed line is the $\SI{190}{\milli\jansky\per\arcsecond\squared\kilo\meter\per\second}$ moment 0 contour. The white dashed line represents the \SI{0.01}{\milli\jansky\per\arcsecond\squared} contour line in the left panel, and the \SI{9}{\kelvin} contour line in the right panel. The gray and white arrows represent the systemic inflow direction.}
   \label{fig:b_1_1_d2}
\end{figure*}%
In the first simulation, we considered a scenario where the disk moves through a highly turbulent medium ($\sigma_\mathrm{turb}=\SI{1}{\kilo\meter\per\second}$) with a large systemic velocity ($v_\mathrm{sys}=\SI{1}{\kilo\meter\per\second}$). Here, the ISM gas is highly energetic, so that only gas in a small volume around the disk remains bound to the star, $R_\mathrm{bound}=\SI{444}{\astronomicalunit}$.

The consequences are two-fold. First, the angular momentum flux through this small volume is the lowest of all simulations and variable on the smallest timescales (see the blue line in Fig. \ref{fig:b_am}). Therefore, any fallback of material does not produce strong arcs. Second, the average linear momentum flux direction also varies strongly and on short timescales. This means that the inflowing gas does not create distinct density enhancements at specific locations, but is smeared out across the volume, preventing the formation of a long-lived and bright BHL accretion tail, as well as streamers that are distinct from this tail.

These considerations serve as the explanation for the majority of features visible in the synthetic observations shown in Fig. \ref{fig:b_1_1_d2}. In scattered light, sensitive to small changes in the density, the most distinct non-disk feature is a straight extended structure above the disk. In addition, one arc that is spatially separate from the main accretion tail can be seen, originating from the right, but only faintly visible. We find that arced structures arise at different times as well, but, due to the high time variability, we find that they persist only on timescales of a few $\si{\kilo\year}$. The flux of all of these structures is low and likely not observable. The CO molecular line emission is not as sensitive to these structures. The moment 1 map only shows the upper structure, which is clearly dynamically separated from the background. We find that this particular structure is not a streamer, but an outflow caused by the interaction with the unbound gas encountering a disk. Apart from a small-scale and short-lived minor infall component on the right side, with no corresponding signal in the scattered light, it is the only visible structure with moment 0 signal beyond \SI{190}{\milli\jansky\per\arcsecond\squared\kilo\meter\per\second}, as indicated by the contour line in Fig. \ref{fig:b_1_1_d2}. The peak brightness temperature of the CO emission of the envelope structures is especially faint; the outflow is barely visible and likely not discernible from the background emission observationally. The infall appears more extended here, but at very low intensity, suggesting that the accretion tail cannot accumulate material due to the frequent and strong velocity fluctuations, and likely could not be detected in observations.

We note that we find substructures in the disk in both scattered light and CO synthetic observations. This is also the case for the scenarios described in subsequent sections, and their discussion is the subject of future work.

\subsubsection{Reduced systemic velocity: Multiple arced streamers}
\begin{figure*}[htp]
    \centering\includegraphics[width=.9\linewidth]{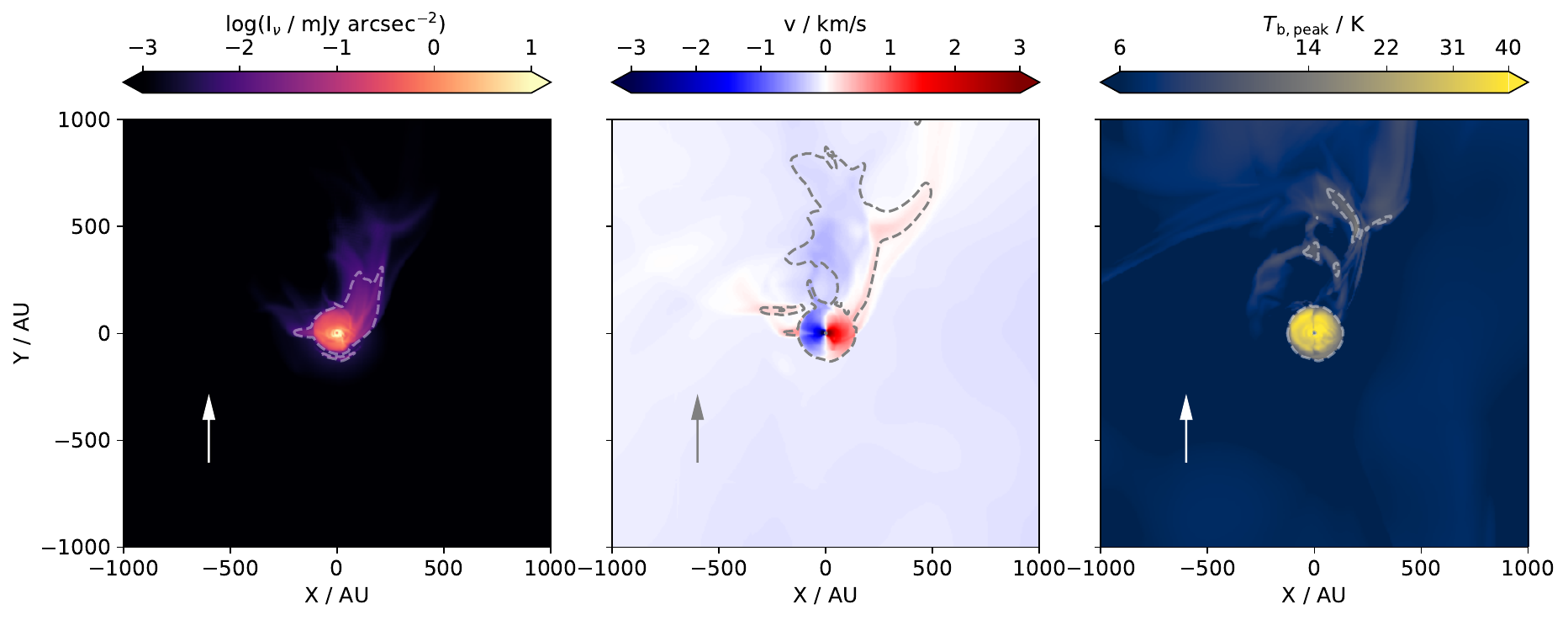}
    \caption{Same as Fig. \ref{fig:b_1_1_d2}, but for simulation number 2, and the level of the moment 0 contour line was adjusted to $\SI{225}{\milli\jansky\per\arcsecond\squared\kilo\meter\per\second}$ to enclose the area of the streamer. Additionally, the color range of the brightness temperature map was adjusted.}
    \label{fig:b_0.5_0.5_d2}
\end{figure*}%
For fallback streamers to arise naturally in a simulation with no initial density enhancement, it is essential that the gas encountering the disk remains bound to the star after the encounter, so that the angular momentum flux is visible as an arc. In simulation number 2, we reduced the systemic velocity by a factor of 2, $v_\mathrm{sys}=\SI{0.5}{\kilo\meter\per\second}$, while keeping the ratio between turbulent velocity dispersion and systemic velocity constant, $\sigma_\mathrm{turb}/v_\mathrm{sys}=1$. As a result, we find $R_\mathrm{bound}=\SI{1775}{\astronomicalunit}$.

We expect that, analogously to the scenario in the previous section, significant variations in the infall direction occur, because the systemic velocity and turbulent velocity dispersion are equal. However, unlike previously, the variations should occur over longer timescales due to the larger $R_\mathrm{bound}$. For the same reason, the angular momentum flux should be higher. Indeed, both expectations are fulfilled in the simulation. The angular momentum flux is up to two orders of magnitude higher than previously (see the purple line in Fig. \ref{fig:b_am}). The direction of the linear momentum flux (see Fig. \ref{fig:b_mom}) and the angular momentum flux vary comparatively less strongly and on longer timescales.

Here, we find that many spatially distinct infall structures are visible in the scattered light images and the CO emission. This is shown in Fig. \ref{fig:b_0.5_0.5_d2}. In the scattered light image, the strongest intensity originates from a streamer originating from the top right with a flux of $\approx\SI{0.01}{\milli\jansky\per\arcsecond\squared}$, while the left structure is less apparent. In the CO moment 1 map, these two streamers are clearly revealed as dynamically separate from each other and the disk. The structures are visible with ${>}\SI{225}{\milli\jansky\per\arcsecond\squared\kilo\meter\per\second}$ across the majority of their spatial extent.

The moment 1 map also reveals that dynamically, the streamers seen in the scattered light consist of multiple sub-streamers: The left one appears to have two components, whereas the right one has three. In the peak brightness temperature map, only the streamer to the right is visible with two components with $T_\mathrm{b,peak}>\SI{9}{\kelvin}$. This brightness temperature is comparable to values observed in the extended structures of GM Tau \citep{huang2021}. We find that the big streamer also persists for the longest time, ${\sim}\SI{15}{\kilo\year}$. The visible part of left streamer is only weakly arced, likely because it was created earlier in the simulation. In fact, the minimum of the angular momentum flux at ${\sim}\SI{15}{\kilo\year}$ seen in Fig. \ref{fig:b_am} seems to separate the formation events leading to the two streamers on either side of the disk, as it also corresponds to a shift in linear momentum direction.

\begin{figure*}[htp]
    \centering\includegraphics[width=\linewidth]{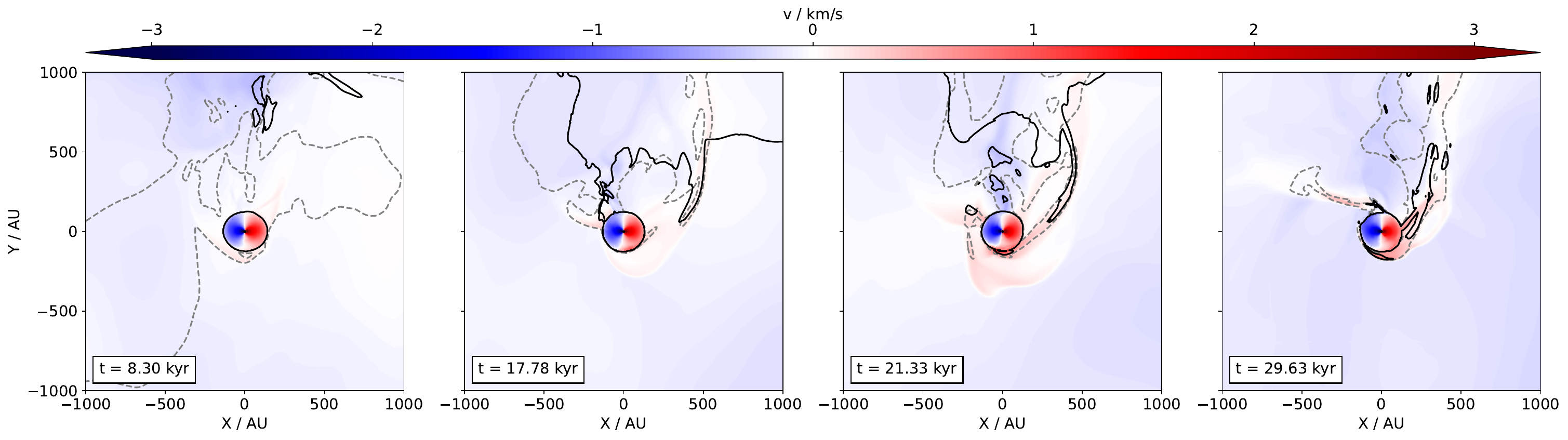}
    \caption{Time series of synthetic CO moment 1 maps as in Fig. \ref{fig:b_0.5_0.5_d2} for the same simulation. The additional black lines are the $\SI{7}{\kelvin}$ contours of the CO peak brightness temperature.}
    \label{fig:b_0.5_0.5_d2_t}
\end{figure*}%
As the structures in this setup are highly time dependent due to the strong turbulence, we show a time series of the CO emission in four simulation snapshots in Fig. \ref{fig:b_0.5_0.5_d2_t}. At first, only weak non-Keplerian features arise around the disk, and the locations of the peak of the CO emission is not correlated with it. In the second panel, the first streamer on the left has formed, and the formation of the one on the right has started. Both big streamers are visible in the third panel, but a third one at the top also appears. It is distinct at this time, but merges with the right streamer by the time of the snapshot shown in Fig. \ref{fig:b_0.5_0.5_d2}. In the end, as seen in the fourth panel, the remaining structures are narrow, fainter, less extended inflows starting to dissipate. In a more realistic scenario with strong turbulence persisting for longer times, continued interactions may continuously lead to the creation of new, strong streamers as old ones dissipate. We therefore expect that fallback streamers could arise abundantly in this type of environment, leading to an increased detection probability compared to the interaction with a single cloudlet.

\begin{figure}[htp]
    \centering\includegraphics[width=.9\linewidth]{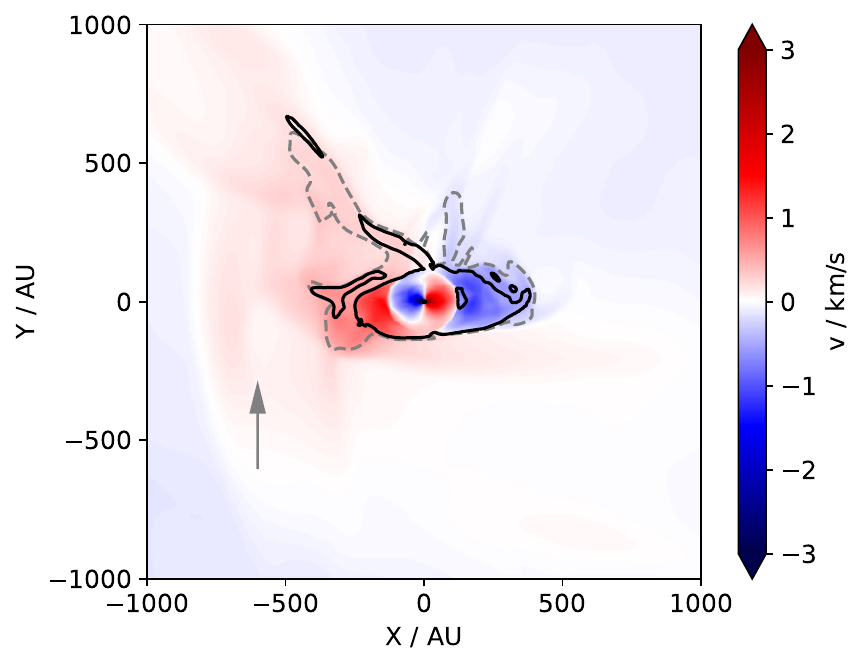}
    \caption{Same as Fig. \ref{fig:b_0.5_0.5_d2}, but for simulation number 3. The level of the moment 0 contour is $\SI{280}{\milli\jansky\per\arcsecond\squared\kilo\meter\per\second}$. The black line shows the $\SI{9}{\kelvin}$ peak brightness temperature contour.}
    \label{fig:b_0.5_0.5_d2_large}
\end{figure}%
This strong turbulence scenario is highly sensitive to the maximal wave number, $k_\mathrm{max}$, of the turbulence power spectrum, that is, the presence or absence of small-scale fluctuations. We explore this sensitivity by reducing this value such that $k_\mathrm{max}=2\pi/(\SI{1000}{\astronomicalunit})$ in simulation number 3. This impacts the arising structures in two major ways, as can be seen in Fig. \ref{fig:b_0.5_0.5_d2_large}. First, the apparent number of subcomponents decreases, and the spatial separation of the two big streamers increases. The streamers and the ISM appear less chaotic in the CO emission. Both streamers are clearly visible in the moment 0 map (signal $>\SI{280}{\milli\jansky\per\arcsecond\squared\kilo\meter\per\second}$) and the peak brightness temperature of the CO molecular emission ($>\SI{9}{\kelvin}$). Combined, it appears that they form a new disk around the existing one, but it can be seen that they originate from two spatially and dynamically distinct mass reservoirs. The formation of these reservoirs is due to two distinct accretion episodes with distinct linear momentum flux. The pink line in Fig. \ref{fig:b_mom} shows a strong change at ${\sim}\SI{20}{\kilo\year}$ that separates the two episodes. In addition, the left mass reservoir is connected to the disk in a second, less arced streamer.

We therefore expect bright streamers with spatially well separated origin, like the connection to extended mass reservoirs, to form in dense (accretion rate $\dot{M}_\mathrm{sys}>\SI{e-8}{\solarmass\per\year}$) environments where small-scale turbulent modes are suppressed or have dissipated due to some physical process.

\subsubsection{Reduced turbulence: Single arced streamer}
\begin{figure*}[htp]
   \centering\includegraphics[width=.9\linewidth]{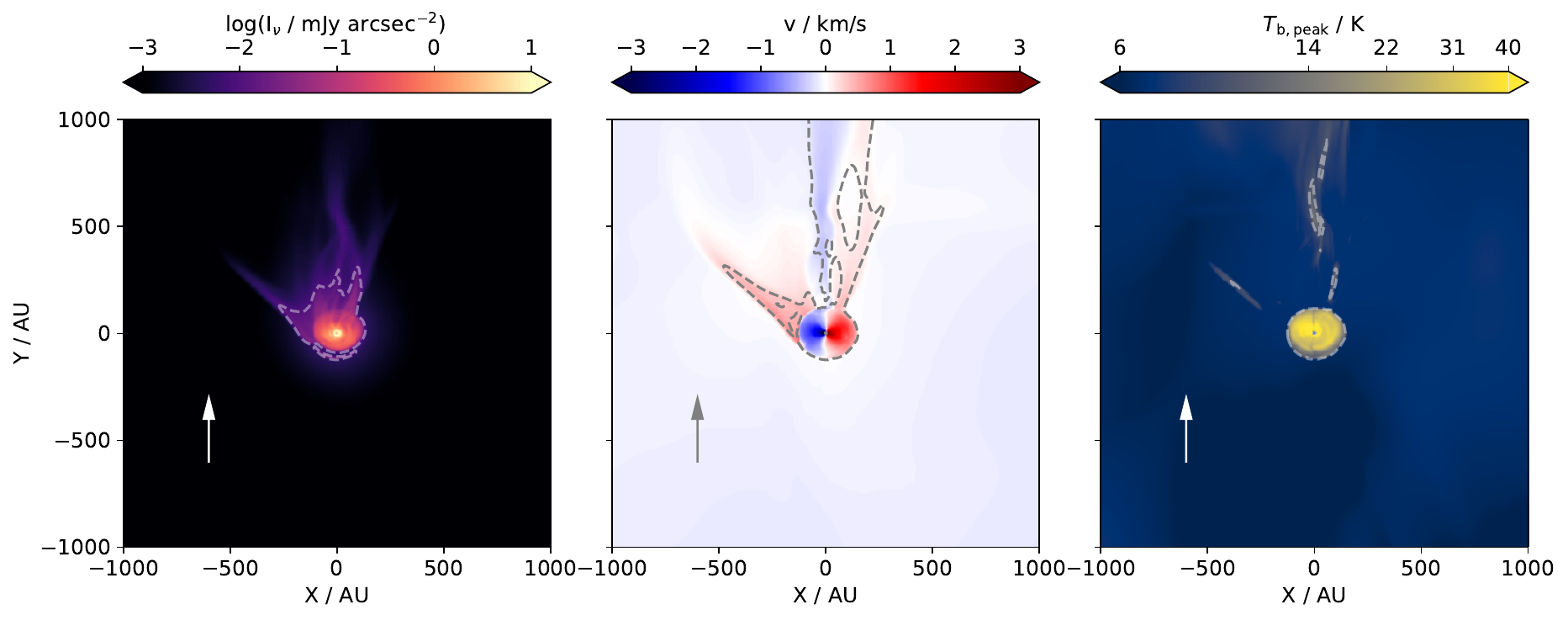}
   \caption{Same as Fig. \ref{fig:b_1_1_d2}, but for simulation number 4. Here, the level of the gray dashed moment 0 contour line is $\SI{200}{\milli\jansky\per\arcsecond\squared\kilo\meter\per\second}$.}
   \label{fig:b_0.5_0.25_d2}
\end{figure*}%
For simulation number 4, we reduced the turbulence amplitude relative to the systemic velocity by a factor of 2, that is, $\sigma_\mathrm{turb}/v_\mathrm{sys}=0.5$. In this scenario, the large-scale turbulent modes are expected to change the direction of inflowing material to a lesser extent. Indeed, the green line in Fig. \ref{fig:b_mom} shows a smaller peak to peak variation of the linear momentum flux direction. It shows a smaller time variability, in part because the turbulent modes with higher $k$ no longer play a significant role. Analogously, the absolute value of the angular momentum flux exhibits less time variability, and is marginally higher in value compared to simulations 2 and 3 at $t>\SI{10}{\kilo\year}$. The limiting radius for bound accretion is $R_\mathrm{bound}=\SI{3155}{\astronomicalunit}$.

The emerging infall structures show two predominant features. One is the previously absent straight inflow originating from the BHL accretion tail, which can now be identified more clearly due to less severe turbulent fluctuations. It is supplied by the flow related to the systemic velocity, which carries no angular momentum relative to the star-disk system. Consequently, the resulting feature is not arced and has a substantially extended morphology, so that the appearance of this accretion tail is less reminiscent of a typical streamer in scattered light and the CO moment maps (see Fig. \ref{fig:b_0.5_0.25_d2}). The intensity of the molecular emission is low, so that the accretion tail is very faint with only some areas with $T_\mathrm{b,peak}>\SI{9}{\kelvin}$, and a distinction from the background in the peak brightness temperature map is difficult.

The second feature is an arc of gas, located at the bottom left edge of the disk. The moment 1 map of the CO emission reveals that it is an inflowing structure with a velocity that is different from the straight accretion tail, as shown in the middle panel of Fig. \ref{fig:b_0.5_0.25_d2}, confirming it as a fallback streamer. Compared to the previously discussed case with higher turbulence, only a single, more strongly arced streamer arises in this scenario. Its spatial origin is located close to the straight accretion tail, which is expected due to the lower amplitude of the turbulent waves. The corresponding CO emission is dim, like the straight accretion tail, but the arc is encompassed by the $\SI{200}{\milli\jansky\per\arcsecond\squared\kilo\meter\per\second}$ moment 0 contour. During earlier times in this simulation, an additional arced streamer is visible for a short time interval, from ${\sim}\SI{20}{\kilo\year}$ to ${\sim}\SI{25}{\kilo\year}$. At this time, the qualitative inflow structure is similar to the bottom and left streamers found in simulation 2 (see Fig. \ref{fig:b_0.5_0.5_d2}), but the bottom streamer is spatially and dynamically close to the accretion tail as a result of the lower turbulence. As a result, it has dissipated at the time shown in Fig. \ref{fig:b_0.5_0.25_d2}.

\begin{figure*}[htp]
   \centering\includegraphics[width=.9\linewidth]{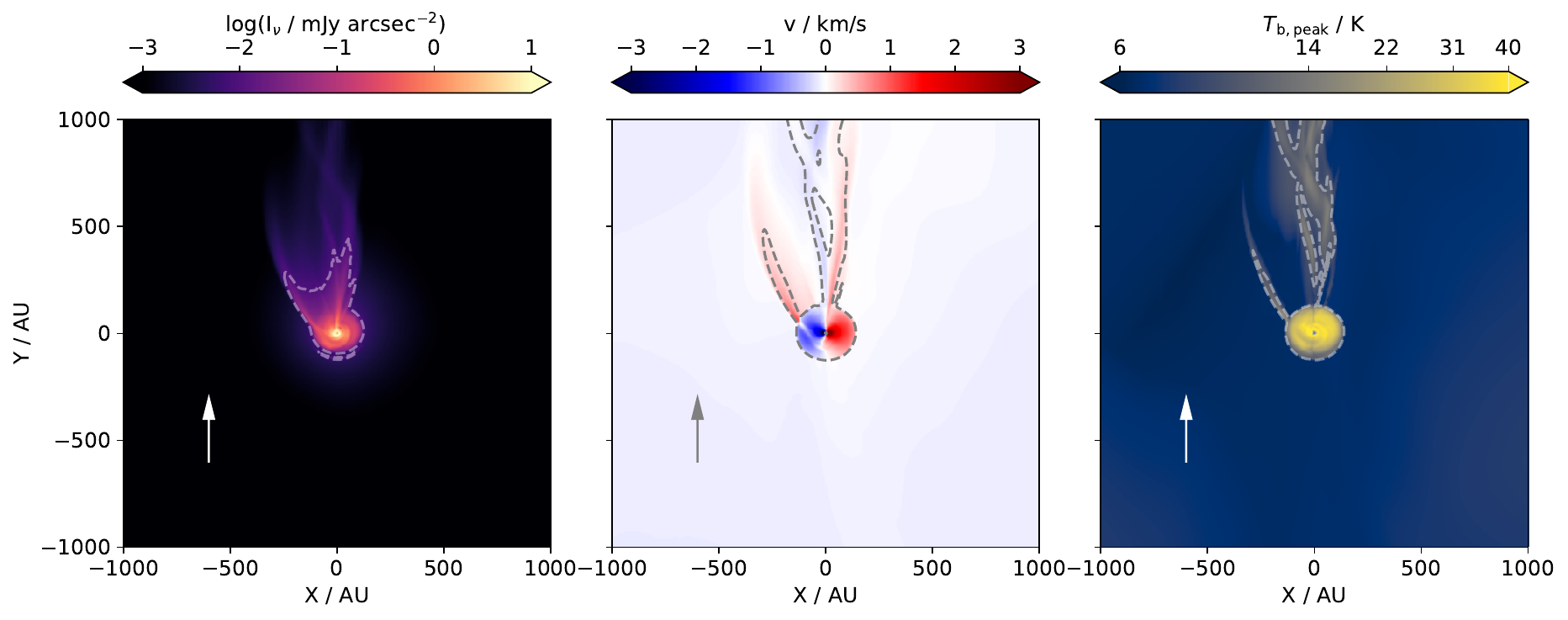}
   \caption{Same as Fig. \ref{fig:b_1_1_d2}, but for simulation number 5, with the level of the moment 0 contour line being $\SI{235}{\milli\jansky\per\arcsecond\squared\kilo\meter\per\second}$.}
   \label{fig:b_0.5_0.1_d2}
\end{figure*}%
Further reduction of the ISM turbulence in simulation 5 further lowers the amplitude of the turbulent velocity fluctuations. To investigate the structures arising in this case, we chose $\sigma_\mathrm{turb}/v_\mathrm{sys}=0.2$. Here, the time variability of the angular momentum and linear momentum fluxes becomes negligible, as indicated by the orange line in Figs. \ref{fig:b_am} and \ref{fig:b_mom}. Due to $R_\mathrm{bound}$ being the highest out of all considered simulations, $R_\mathrm{bound}=\SI{4930}{\astronomicalunit}$, the absolute value of the angular momentum flux is also maximal. The resulting structures are shown in Fig. \ref{fig:b_0.5_0.1_d2}.

We find that, under these conditions, almost all gas inflowing onto the disk from the environment is part of the straight accretion tail. Scattered light and CO moment maps also show a single, strongly arced fallback-streamer, which is expected from the high angular momentum flux. Its signal is ${>}\SI{235}{\milli\jansky\per\arcsecond\squared\kilo\meter\per\second}$ in moment 0. Due to the low time variability, both the accretion tail and the arced streamer can be seen more clearly in the peak brightness temperature, with $T_\mathrm{b,peak}>\SI{9}{\kelvin}$. While no additional streamers emerge at any time during the simulation, this result indicates that even low levels of turbulence can lead to a high angular momentum flux relative to the disk, forming an arced streamer in the process.

\section{Discussion}\label{sec:discussion}
\subsection{Fallback streamers should be common}
Previous modeling of streamers that only consider them as originating from clear mass filaments, or as the signatures of the accretion of massive gas cloudlets (see Section \ref{sec:res_cloudlet}), implies that streamers should be rare. This conclusion is strengthened further by the fact that streamer-like structures do not persist for a large fraction of the disk lifetime: only ${\sim}1\%$ of it assuming a conservative disk lifetime of \SI{1}{\mega\year}.

Unless such encounters are very frequent to replenish the inflow and the related signatures, they fall short of explaining the high abundance of late-infall streamers suggested by observations \citep{gupta2023,garufi2024}. In this case, the fallback streamers we find in the described simulations may offer an explanation. Here, streamers are a common phenomenon for disk-star systems moving through a dense environment. Even though individual streamers do not last considerably longer than in the case of a cloudlet encounter, their emergence does not depend on a specific encounter, but are the result of smaller asymmetries caused by turbulence. As a result, they should arise continuously, increasing the total time they may be observable. In fact, even a low level of turbulence can create a single arced streamer on top of the straight accretion tail.

The fallback streamers, even though their morphology is similar to a stream of material originating from a mass reservoir, do not hold the same information about the large-scale environment. In particular, the infall direction that could be obtained through dynamical modeling is not related to the spatial location of an over-density in the molecular cloud, and a direct connection may not be found. Even in the simplified case of cloudlet accretion, the fallback streamers origin is different from the origin of the cloudlet initially encountering the disk. For turbulent accretion, especially in a highly turbulent environment, the orientation of the streamers is fully random. Here, only the straight infall directly from the accretion tail can be used to determine the systemic velocity of the disk.

Infall of material via BHL accretion after the initial core collapse is also found in parsec-scale star-formation simulations \citep{padoan2025,pelkonen2025}. Previous studies attribute various protoplanetary disk features to this phenomenon \citep{hennebelle2017,kuffmeier2018,kuffmeier2023,winter2024}, further emphasizing its relevance and the need to more in-depth studies of its implications. In this work, we considered the occurrence and morphology of streamers arising in the BHL accretion scenario, and we explored a multitude of environmental conditions. Such considerations were generally not subject of previous large-scale simulations. We focussed on the environment around a single star with a simplified treatment of the environment, identifying fallback as the source for the streamers in our simulations. In the future, our findings should be compared to the morphology and occurrence of accretion structures around disks arising self-consistently in parsec-scale star-formation simulations. The majority of such simulations do not currently reach far into the Class II phase, though.

\subsection{Streamer morphology as a constraint for local cloud conditions}
The quantity and morphology of streamers depends on the local conditions of the ISM as the system is moving through it, especially on the turbulence strength. The observational detection and characterization of streamers onto Class II disks could therefore be used to determine the conditions of their environment. If multiple, arced streamers are detected simultaneously, the disk environment may be very turbulent, whereas the presence of a bright BHL accretion tail suggests lower levels of turbulence. In either case, these structures would indicate a substantial infall rate onto the disk, $\dot{M}_\mathrm{sys}\sim\SI{e-8}{\solarmass\per\year}$.

The detection of a straight inflow signature, which can be identified as the accretion tail caused by the systemic flow, serves as strong evidence for a dense environment that may cause additional inflow signatures. The orientation of the straight accretion tail should then be correlated with the proper motion of the disk's host star determined by, for example, Gaia. In contrast, the orientation of the fallback streamers caused by turbulence should be uncorrelated. Detecting a single streamer would be indicative of a low turbulence environment, but if no corresponding BHL accretion tail is found, a formation scenario related to the cloudlet case may be the favored explanation, implying that the environmental density is low. It may then be caused by such a single encounter, or originate from a mass reservoir. We note that in environments with moderate levels of turbulence, like in our simulation number 4, the presence of an arced streamer may be missed by observations as emission is faint.

The spatial separation and multiplicity of streamers can provide information on the spatial scales the turbulence operates on. For high separation, but low multiplicity that is ${>}1$, it might be indicative that small-scale turbulence may be suppressed by some physical process, whose origin may shed more light on the conditions present in the environment of Class II disks. On the other hand, if a high spatial separation is accompanied by a high multiplicity, small-scale turbulence may be operating.

\subsection{Implications for disk dynamics and planet formation}
The high abundance of observational signatures of streamers, which can be directly related to asymmetric inflow from a dense, large-scale environment engulfing the disk, has strong implications for the dynamical evolution of Class II protoplanetary disks, and thereby for planet formation. Asymmetric inflow has been found to cause a torque that can facilitate the formation of a vortex \citep{bae2015,kuznetsova2022}. This vortex could act as a trap for solids, enhancing the local solid-to-gas ratio to aid in the formation of planetesimals. Furthermore, the inflowing gas itself may locally enhance the turbulence \citep{lesur2015,hennebelle2017} and related angular momentum transport in the disk, at multiple locations in some cases, which could cause accretion outbursts and allow for the formation of pressure bumps at multiple locations. Gravitational instability may also be triggered as a result of inflow \citep{kl2016,kuffmeier2018}. We will investigate some of these aspects in future work.

In addition to impacting disk dynamics, the late infall that causes these streamers can alleviate the dust-mass-budget problem in Class II disks. Another possible solution for this problem is the early onset of planet formation during the Class 0/I phase, which can strongly profit from infall \citep{drazkowska2018,morbidelli2022,huehn2025}, though it remains unclear whether the dynamical conditions in these young disks are favorable for planetesimal formation \citep{carrera2025}. If dynamically more quiescent Class II disks, however, can also undergo significant accretion over extended time periods, planet formation could start at these late stages and still have enough material to form planets of the observed masses, perhaps even forming planets in multiple generations.

\subsection{Model caveats}
We modeled the BHL accretion flow onto a protoplanetary disk moving through the ISM in a simplified way. In addition to the described approximation introduced by not driving the turbulence, but just introducing it as an initial condition, we did not model heating and cooling of the gas. Shock heating at the impact locations, where the inflowing gas streamers hit the disk, may be a significant heating source that is also not captured by the Monte-Carlo temperature calculation we perform during post-processing. However, we do not expect a qualitative change to the streamer morphology by a more detailed temperature treatment, but rather an effect on the disk chemistry and the detailed morphology of potentially arising accretion spirals \citep{vangelder2021}. Other types of modeling may be more suitable for detailed treatment of the impact zone. Furthermore, we did not consider the influence of magnetic fields, which may affect the shape and the angular momentum of the accreted gas, and we neglect stellar feedback to the accretion flow.

When computing the synthetic images, the freeze-out and photodissociation of CO was neglected. The freeze-out of CO onto interstellar dust is particularly relevant for the cold regions far from stellar heating sources. It could reduce the background emission, which would enhance the contrast of the streamer structures and thereby their observability. Conversely, the photodissociation of CO could reduce emission close to the star, balanced by shielding effects of molecular hydrogen and CO itself. To estimate the strength of this effect, we performed order-of-magnitude calculations of the CO column density, assuming photodissociation as the only reaction. We assumed a moderately high unattenuated UV field strength $\chi=\num{e5}$, given in units of the Draine field and at a reference radius of \SI{100}{\astronomicalunit}, used reaction rates from the UMIST database for astrochemistry \citep{woodall2007}, and employed the self-shielding prescription from \citet{woitke2009}. We neglected extinction by the ISM dust. In regions with a lower density of \SI{e-19}{\gram\per\centi\meter\cubed}, 50\% of CO at a distance of \SI{1000}{\astronomicalunit} may dissociate over a streamer lifetime of \SI{10}{\kilo\year}, but almost all CO inside \SI{500}{\astronomicalunit} could be dissociated. In regions with a high density of \SI{e-17}{\gram\per\centi\meter\cubed}, only 2\% of CO is dissociated close to the star at \SI{200}{\astronomicalunit}. Therefore, photodissociation may play a significant role for low-density streamers arising in high-turbulence regions, but more detailed modeling would be required to accurately assess how astrochemical processes affect the observability of the streamers we find using CO lines.

\section{Conclusions}\label{sec:conclusions}
We performed 3D hydrodynamical simulations to investigate the formation of late-infall streamers in dense environments. We modeled late infall as an encounter with a gas cloudlet on an inclined orbit, and we considered the emergence of streamers while a protoplanetary disks moves through the ISM subject to different levels of turbulence. For the latter, a summary of the resulting streamers in a single channel is shown in Appendix \ref{app:channels}. We draw the following conclusions.
\begin{itemize}
    \item The encounter of a Class II disk with an expanding cloudlet results in the formation of a "fallback" streamer, where material that initially overshot remains bound and falls back onto the disk with relative angular momentum.
    \item The apparent spatial origin of the cloudlet streamer differs from the original origin of the cloudlet. Consequently, the continued inflow from the original direction shocks material that is falling back, enhancing the streamer.
    \item A cloudlet encounter leads to the formation of single, strongly arced streamer and may be relevant in environments with otherwise low density. It dissipates after ${\sim}\SI{10}{\kilo\year}$. Despite its bright CO emission, the short lifetime makes it unlikely to observe cloudlet streamers unless the frequency of such encounters is high.
    \item Streamers with different morphologies and multiplicities can arise naturally given a high enough accretion rate (${\sim}\SI{e-8}{\solarmass\per\year}$), as bound material accumulated during BHL accretion falls back onto the disk with relative angular momentum, similarly to a cloudlet encounter.
    \item If the disk has a high systemic velocity, $v_\mathrm{sys}=\SI{1}{\kilo\meter\per\second}$, a majority of accreted material does not remain bound to the disk, so that no extended streamers or even outflows arise.
    \item For $v_\mathrm{sys}=\SI{0.5}{\kilo\meter\per\second}$ and a turbulent velocity dispersion of $\sigma_\mathrm{turb}=\SI{0.5}{\kilo\meter\per\second}$, accretion is chaotic, but a substantial amount of gas remains bound. Here, multiple arced streamers with subcomponents emerge, and a BHL accretion tail cannot be seen.
    \item In the absence of small-scale turbulent modes, two streamers connected to distinct mass reservoirs arise, suggesting that these modes may be suppressed in reality if such streamers are observed.
    \item Reducing the turbulence to $\sigma_\mathrm{turb}=\SI{0.25}{\kilo\meter\per\second}$ or $\sigma_\mathrm{turb}=\SI{0.1}{\kilo\meter\per\second}$ allows for the formation of a single arced streamer and a straight BHL tail. Secondary streamers may arise briefly on the timescale of a few \si{\kilo\year}.
\end{itemize}
Our findings suggest that while late infall with a clear origin, represented by cloudlet encounters, can create streamers, they can also arise naturally as an evolved protoplanetary disk moves through the turbulent ISM. This can reconcile ubiquitous detections of late-infall streamers with their short lifetime. Because their morphology is sensitive to the turbulent conditions of the environment and the accretion rate onto the disk, detailed studies of streamers may shed light on the conditions present in the environment of planet-forming disks, which may help to improve our understanding of planet formation.

\begin{acknowledgements}
We thank Jiahan Shi, Jorge Pérez González, and Dmitry Semenov for insightful discussions. L.-A. H. acknowledges funding by the DFG via the Heidelberg Cluster of Excellence STRUCTURES in the framework of Germany's Excellence Strategy (grant EXC-2181/1 -- 390900948). We acknowledge support by the High Performance and Cloud Computing Group at the Zentrum für Datenverarbeitung of the University of Tübingen, the state of Baden-Württemberg through bwHPC and the German Research Foundation (DFG) through grants INST 35/1134-1 FUGG, 35/1597-1 FUGG and 37/935-1 FUGG.
\end{acknowledgements}
\bibliography{references}

\begin{thebibliography}{54}
\expandafter\ifx\csname natexlab\endcsname\relax\def\natexlab#1{#1}\fi

\bibitem[{{Bae} {et~al.}(2015){Bae}, {Hartmann}, \& {Zhu}}]{bae2015}
{Bae}, J., {Hartmann}, L., \& {Zhu}, Z. 2015, \apj, 805, 15

\bibitem[{{Ben{\'\i}tez-Llambay} \& {Masset}(2016)}]{benitez-llambay2016}
{Ben{\'\i}tez-Llambay}, P. \& {Masset}, F.~S. 2016, \apjs, 223, 11

\bibitem[{{Bondi}(1945)}]{bondi1945}
{Bondi}, A. 1945, Journal of Applied Physics, 16, 539

\bibitem[{{Calcino} {et~al.}(2025){Calcino}, {Price}, {Hilder}, {Christiaens}, {Speedie}, \& {Ormel}}]{calcino2025}
{Calcino}, J., {Price}, D.~J., {Hilder}, T., {et~al.} 2025, \mnras, 537, 2695

\bibitem[{{Carrera} {et~al.}(2025){Carrera}, {Davenport}, {Simon}, {Baehr}, {Birnstiel}, {Hall}, {Rea}, \& {Stammler}}]{carrera2025}
{Carrera}, D., {Davenport}, A., {Simon}, J.~B., {et~al.} 2025, \apj, 990, 39

\bibitem[{{Dai} {et~al.}(2015){Dai}, {Facchini}, {Clarke}, \& {Haworth}}]{dai2015}
{Dai}, F., {Facchini}, S., {Clarke}, C.~J., \& {Haworth}, T.~J. 2015, \mnras, 449, 1996

\bibitem[{{Dominik} {et~al.}(2021){Dominik}, {Min}, \& {Tazaki}}]{dominik2021}
{Dominik}, C., {Min}, M., \& {Tazaki}, R. 2021, {OpTool: Command-line driven tool for creating complex dust opacities}, Astrophysics Source Code Library, record ascl:2104.010

\bibitem[{{Dr{\k{a}}{\.z}kowska} {et~al.}(2023){Dr{\k{a}}{\.z}kowska}, {Bitsch}, {Lambrechts}, {Mulders}, {Harsono}, {Vazan}, {Liu}, {Ormel}, {Kretke}, \& {Morbidelli}}]{drazkowska2023}
{Dr{\k{a}}{\.z}kowska}, J., {Bitsch}, B., {Lambrechts}, M., {et~al.} 2023, in Astronomical Society of the Pacific Conference Series, Vol. 534, Protostars and Planets VII, ed. S.~{Inutsuka}, Y.~{Aikawa}, T.~{Muto}, K.~{Tomida}, \& M.~{Tamura}, 717

\bibitem[{{Dr{\k{a}}{\.z}kowska} \& {Dullemond}(2018)}]{drazkowska2018}
{Dr{\k{a}}{\.z}kowska}, J. \& {Dullemond}, C.~P. 2018, \aap, 614, A62

\bibitem[{{Dubinski} {et~al.}(1995){Dubinski}, {Narayan}, \& {Phillips}}]{dubinski1995}
{Dubinski}, J., {Narayan}, R., \& {Phillips}, T.~G. 1995, \apj, 448, 226

\bibitem[{{Dullemond} {et~al.}(2012){Dullemond}, {Juhasz}, {Pohl}, {Sereshti}, {Shetty}, {Peters}, {Commercon}, \& {Flock}}]{radmc3d}
{Dullemond}, C.~P., {Juhasz}, A., {Pohl}, A., {et~al.} 2012, {RADMC-3D: A multi-purpose radiative transfer tool}, Astrophysics Source Code Library, record ascl:1202.015

\bibitem[{{Dullemond} {et~al.}(2022){Dullemond}, {Kimmig}, \& {Zanazzi}}]{dullemond2022}
{Dullemond}, C.~P., {Kimmig}, C.~N., \& {Zanazzi}, J.~J. 2022, \mnras, 511, 2925

\bibitem[{{Dullemond} {et~al.}(2019){Dullemond}, {K{\"u}ffmeier}, {Goicovic}, {Fukagawa}, {Oehl}, \& {Kramer}}]{dullemond2019}
{Dullemond}, C.~P., {K{\"u}ffmeier}, M., {Goicovic}, F., {et~al.} 2019, \aap, 628, A20

\bibitem[{{Falgarone} {et~al.}(2004){Falgarone}, {Hily-Blant}, \& {Levrier}}]{falgarone2004}
{Falgarone}, E., {Hily-Blant}, P., \& {Levrier}, F. 2004, \apss, 292, 89

\bibitem[{{Garufi} {et~al.}(2024){Garufi}, {Ginski}, {van Holstein}, {Benisty}, {Manara}, {P{\'e}rez}, {Pinilla}, {Ribas}, {Weber}, {Williams}, {Cieza}, {Dominik}, {Facchini}, {Huang}, {Zurlo}, {Bae}, {Hagelberg}, {Henning}, {Hogerheijde}, {Janson}, {M{\'e}nard}, {Messina}, {Meyer}, {Pinte}, {Quanz}, {Rigliaco}, {Roccatagliata}, {Schmid}, {Szul{\'a}gyi}, {van Boekel}, {Wahhaj}, {Antichi}, {Baruffolo}, \& {Moulin}}]{garufi2024}
{Garufi}, A., {Ginski}, C., {van Holstein}, R.~G., {et~al.} 2024, \aap, 685, A53

\bibitem[{{Garufi} {et~al.}(2022){Garufi}, {Podio}, {Codella}, {Segura-Cox}, {Vander Donckt}, {Mercimek}, {Bacciotti}, {Fedele}, {Kasper}, {Pineda}, {Humphreys}, \& {Testi}}]{garufi2022}
{Garufi}, A., {Podio}, L., {Codella}, C., {et~al.} 2022, \aap, 658, A104

\bibitem[{{Ginski} {et~al.}(2021){Ginski}, {Facchini}, {Huang}, {Benisty}, {Vaendel}, {Stapper}, {Dominik}, {Bae}, {M{\'e}nard}, {Muro-Arena}, {Hogerheijde}, {McClure}, {van Holstein}, {Birnstiel}, {Boehler}, {Bohn}, {Flock}, {Mamajek}, {Manara}, {Pinilla}, {Pinte}, \& {Ribas}}]{ginski2021}
{Ginski}, C., {Facchini}, S., {Huang}, J., {et~al.} 2021, \apjl, 908, L25

\bibitem[{{Gupta} {et~al.}(2023){Gupta}, {Miotello}, {Manara}, {Williams}, {Facchini}, {Beccari}, {Birnstiel}, {Ginski}, {Hacar}, {K{\"u}ffmeier}, {Testi}, {Tychoniec}, \& {Yen}}]{gupta2023}
{Gupta}, A., {Miotello}, A., {Manara}, C.~F., {et~al.} 2023, \aap, 670, L8

\bibitem[{{Gupta} {et~al.}(2024){Gupta}, {Miotello}, {Williams}, {Birnstiel}, {Kuffmeier}, \& {Yen}}]{gupta2024}
{Gupta}, A., {Miotello}, A., {Williams}, J.~P., {et~al.} 2024, \aap, 683, A133

\bibitem[{{Hanawa} {et~al.}(2024){Hanawa}, {Garufi}, {Podio}, {Codella}, \& {Segura-Cox}}]{hanawa2024}
{Hanawa}, T., {Garufi}, A., {Podio}, L., {Codella}, C., \& {Segura-Cox}, D. 2024, \mnras, 528, 6581

\bibitem[{{Hennebelle} {et~al.}(2017){Hennebelle}, {Lesur}, \& {Fromang}}]{hennebelle2017}
{Hennebelle}, P., {Lesur}, G., \& {Fromang}, S. 2017, \aap, 599, A86

\bibitem[{{Herczeg} {et~al.}(2023){Herczeg}, {Chen}, {Donati}, {Dupree}, {Walter}, {Hillenbrand}, {Johns-Krull}, {Manara}, {G{\"u}nther}, {Fang}, {Schneider}, {Valenti}, {Alencar}, {Venuti}, {Alcal{\'a}}, {Frasca}, {Arulanantham}, {Linsky}, {Bouvier}, {Brickhouse}, {Calvet}, {Espaillat}, {Campbell-White}, {Carpenter}, {Chang}, {Cruz}, {Dahm}, {Eisl{\"o}ffel}, {Edwards}, {Fischer}, {Guo}, {Henning}, {Ji}, {Jose}, {Kastner}, {Launhardt}, {Principe}, {Robinson}, {Serna}, {Siwak}, {Sterzik}, \& {Takasao}}]{herczeg2023}
{Herczeg}, G.~J., {Chen}, Y., {Donati}, J.-F., {et~al.} 2023, \apj, 956, 102

\bibitem[{{Hoyle} \& {Lyttleton}(1939)}]{hl1939}
{Hoyle}, F. \& {Lyttleton}, R.~A. 1939, Proceedings of the Cambridge Philosophical Society, 35, 405

\bibitem[{{Huang} {et~al.}(2021){Huang}, {Bergin}, {{\"O}berg}, {Andrews}, {Teague}, {Law}, {Kalas}, {Aikawa}, {Bae}, {Bergner}, {Booth}, {Bosman}, {Calahan}, {Cataldi}, {Cleeves}, {Czekala}, {Ilee}, {Le Gal}, {Guzm{\'a}n}, {Long}, {Loomis}, {M{\'e}nard}, {Nomura}, {Qi}, {Schwarz}, {Tsukagoshi}, {van't Hoff}, {Walsh}, {Wilner}, {Yamato}, \& {Zhang}}]{huang2021}
{Huang}, J., {Bergin}, E.~A., {{\"O}berg}, K.~I., {et~al.} 2021, \apjs, 257, 19

\bibitem[{{H{\"u}hn} {et~al.}(2025){H{\"u}hn}, {Dullemond}, {Lebreuilly}, {Klessen}, {Maury}, {Rosotti}, {Hennebelle}, {Pacetti}, {Testi}, \& {Molinari}}]{huehn2025}
{H{\"u}hn}, L.~A., {Dullemond}, C.~P., {Lebreuilly}, U., {et~al.} 2025, \aap, 696, A162

\bibitem[{{Klessen} \& {Hennebelle}(2010)}]{kh2010}
{Klessen}, R.~S. \& {Hennebelle}, P. 2010, \aap, 520, A17

\bibitem[{{Kratter} \& {Lodato}(2016)}]{kl2016}
{Kratter}, K. \& {Lodato}, G. 2016, \araa, 54, 271

\bibitem[{{Krumholz} {et~al.}(2006){Krumholz}, {McKee}, \& {Klein}}]{krumholz2006}
{Krumholz}, M.~R., {McKee}, C.~F., \& {Klein}, R.~I. 2006, \apj, 638, 369

\bibitem[{{Kuffmeier} {et~al.}(2021){Kuffmeier}, {Dullemond}, {Reissl}, \& {Goicovic}}]{kuffmeier2021}
{Kuffmeier}, M., {Dullemond}, C.~P., {Reissl}, S., \& {Goicovic}, F.~G. 2021, \aap, 656, A161

\bibitem[{{Kuffmeier} {et~al.}(2018){Kuffmeier}, {Frimann}, {Jensen}, \& {Haugb{\o}lle}}]{kuffmeier2018}
{Kuffmeier}, M., {Frimann}, S., {Jensen}, S.~S., \& {Haugb{\o}lle}, T. 2018, \mnras, 475, 2642

\bibitem[{{Kuffmeier} {et~al.}(2020){Kuffmeier}, {Goicovic}, \& {Dullemond}}]{kuffmeier2020}
{Kuffmeier}, M., {Goicovic}, F.~G., \& {Dullemond}, C.~P. 2020, \aap, 633, A3

\bibitem[{{Kuffmeier} {et~al.}(2023){Kuffmeier}, {Jensen}, \& {Haugb{\o}lle}}]{kuffmeier2023}
{Kuffmeier}, M., {Jensen}, S.~S., \& {Haugb{\o}lle}, T. 2023, European Physical Journal Plus, 138, 272

\bibitem[{{Kurtovic} {et~al.}(2018){Kurtovic}, {P{\'e}rez}, {Benisty}, {Zhu}, {Zhang}, {Huang}, {Andrews}, {Dullemond}, {Isella}, {Bai}, {Carpenter}, {Guzm{\'a}n}, {Ricci}, \& {Wilner}}]{kurtovic2018}
{Kurtovic}, N.~T., {P{\'e}rez}, L.~M., {Benisty}, M., {et~al.} 2018, \apjl, 869, L44

\bibitem[{{Kuznetsova} {et~al.}(2022){Kuznetsova}, {Bae}, {Hartmann}, \& {Mac Low}}]{kuznetsova2022}
{Kuznetsova}, A., {Bae}, J., {Hartmann}, L., \& {Mac Low}, M.-M. 2022, \apj, 928, 92

\bibitem[{{Lesur} {et~al.}(2015){Lesur}, {Hennebelle}, \& {Fromang}}]{lesur2015}
{Lesur}, G., {Hennebelle}, P., \& {Fromang}, S. 2015, \aap, 582, L9

\bibitem[{{Manara} {et~al.}(2018){Manara}, {Morbidelli}, \& {Guillot}}]{manara2018}
{Manara}, C.~F., {Morbidelli}, A., \& {Guillot}, T. 2018, \aap, 618, L3

\bibitem[{{Masset}(2000)}]{masset2000}
{Masset}, F. 2000, \aaps, 141, 165

\bibitem[{{Morbidelli} {et~al.}(2022){Morbidelli}, {Bailli{\'e}}, {Batygin}, {Charnoz}, {Guillot}, {Rubie}, \& {Kleine}}]{morbidelli2022}
{Morbidelli}, A., {Bailli{\'e}}, K., {Batygin}, K., {et~al.} 2022, Nature Astronomy, 6, 72

\bibitem[{{Padoan} {et~al.}(2025){Padoan}, {Pan}, {Pelkonen}, {Haugb{\o}lle}, \& {Nordlund}}]{padoan2025}
{Padoan}, P., {Pan}, L., {Pelkonen}, V.-M., {Haugb{\o}lle}, T., \& {Nordlund}, {\r{A}}. 2025, Nature Astronomy

\bibitem[{{Palmeirim} {et~al.}(2013){Palmeirim}, {Andr{\'e}}, {Kirk}, {Ward-Thompson}, {Arzoumanian}, {K{\"o}nyves}, {Didelon}, {Schneider}, {Benedettini}, {Bontemps}, {Di Francesco}, {Elia}, {Griffin}, {Hennemann}, {Hill}, {Martin}, {Men'shchikov}, {Molinari}, {Motte}, {Nguyen Luong}, {Nutter}, {Peretto}, {Pezzuto}, {Roy}, {Rygl}, {Spinoglio}, \& {White}}]{palmeirim2013}
{Palmeirim}, P., {Andr{\'e}}, P., {Kirk}, J., {et~al.} 2013, \aap, 550, A38

\bibitem[{{Pelkonen} {et~al.}(2025){Pelkonen}, {Padoan}, {Juvela}, {Haugb{\o}lle}, \& {Nordlund}}]{pelkonen2025}
{Pelkonen}, V.~M., {Padoan}, P., {Juvela}, M., {Haugb{\o}lle}, T., \& {Nordlund}, {\r{A}}. 2025, \aap, 694, A327

\bibitem[{{Pfalzner} {et~al.}(2024){Pfalzner}, {Govind}, \& {Portegies Zwart}}]{pfalzner2024}
{Pfalzner}, S., {Govind}, A., \& {Portegies Zwart}, S. 2024, Nature Astronomy, 8, 1380

\bibitem[{{Pineda} {et~al.}(2023){Pineda}, {Arzoumanian}, {Andre}, {Friesen}, {Zavagno}, {Clarke}, {Inoue}, {Chen}, {Lee}, {Soler}, \& {Kuffmeier}}]{pineda2023}
{Pineda}, J.~E., {Arzoumanian}, D., {Andre}, P., {et~al.} 2023, in Astronomical Society of the Pacific Conference Series, Vol. 534, Protostars and Planets VII, ed. S.~{Inutsuka}, Y.~{Aikawa}, T.~{Muto}, K.~{Tomida}, \& M.~{Tamura}, 233

\bibitem[{{Pineda} {et~al.}(2010){Pineda}, {Goldsmith}, {Chapman}, {Snell}, {Li}, {Cambr{\'e}sy}, \& {Brunt}}]{pineda2010}
{Pineda}, J.~L., {Goldsmith}, P.~F., {Chapman}, N., {et~al.} 2010, \apj, 721, 686

\bibitem[{{Seifried} {et~al.}(2013){Seifried}, {Banerjee}, {Pudritz}, \& {Klessen}}]{seifried2013}
{Seifried}, D., {Banerjee}, R., {Pudritz}, R.~E., \& {Klessen}, R.~S. 2013, \mnras, 432, 3320

\bibitem[{{Shakura} \& {Sunyaev}(1973)}]{ss1973}
{Shakura}, N.~I. \& {Sunyaev}, R.~A. 1973, \aap, 24, 337

\bibitem[{{Speedie} {et~al.}(2024){Speedie}, {Dong}, {Hall}, {Longarini}, {Veronesi}, {Paneque-Carre{\~n}o}, {Lodato}, {Tang}, {Teague}, \& {Hashimoto}}]{speedie2024}
{Speedie}, J., {Dong}, R., {Hall}, C., {et~al.} 2024, \nat, 633, 58

\bibitem[{{Speedie} {et~al.}(2025){Speedie}, {Dong}, {Teague}, {Segura-Cox}, {Pineda}, {Calcino}, {Longarini}, {Hall}, {Tang}, {Hashimoto}, {Paneque-Carre{\~n}o}, {Lodato}, \& {Veronesi}}]{speedie2025}
{Speedie}, J., {Dong}, R., {Teague}, R., {et~al.} 2025, \apjl, 981, L30

\bibitem[{{van Gelder} {et~al.}(2021){van Gelder}, {Tabone}, {van Dishoeck}, \& {Godard}}]{vangelder2021}
{van Gelder}, M.~L., {Tabone}, B., {van Dishoeck}, E.~F., \& {Godard}, B. 2021, \aap, 653, A159

\bibitem[{{Villenave} {et~al.}(2024){Villenave}, {Stapelfeldt}, {Duch{\^e}ne}, {M{\'e}nard}, {Wolff}, {Perrin}, {Pinte}, {Tazaki}, \& {Padgett}}]{villenave2024}
{Villenave}, M., {Stapelfeldt}, K.~R., {Duch{\^e}ne}, G., {et~al.} 2024, \apj, 961, 95

\bibitem[{{Winter} {et~al.}(2024){Winter}, {Benisty}, \& {Andrews}}]{winter2024}
{Winter}, A.~J., {Benisty}, M., \& {Andrews}, S.~M. 2024, \apjl, 972, L9

\bibitem[{{Woitke} {et~al.}(2009){Woitke}, {Kamp}, \& {Thi}}]{woitke2009}
{Woitke}, P., {Kamp}, I., \& {Thi}, W.~F. 2009, \aap, 501, 383

\bibitem[{{Woitke} {et~al.}(2016){Woitke}, {Min}, {Pinte}, {Thi}, {Kamp}, {Rab}, {Anthonioz}, {Antonellini}, {Baldovin-Saavedra}, {Carmona}, {Dominik}, {Dionatos}, {Greaves}, {G{\"u}del}, {Ilee}, {Liebhart}, {M{\'e}nard}, {Rigon}, {Waters}, {Aresu}, {Meijerink}, \& {Spaans}}]{woitke2016}
{Woitke}, P., {Min}, M., {Pinte}, C., {et~al.} 2016, \aap, 586, A103

\bibitem[{{Woodall} {et~al.}(2007){Woodall}, {Ag{\'u}ndez}, {Markwick-Kemper}, \& {Millar}}]{woodall2007}
{Woodall}, J., {Ag{\'u}ndez}, M., {Markwick-Kemper}, A.~J., \& {Millar}, T.~J. 2007, \aap, 466, 1197

\end{thebibliography}
\begin{appendix}
\onecolumn
\section{Turbulent velocity field}\label{app:turbfield}
\begin{figure*}[htp]
    \centering\includegraphics[width=\linewidth]{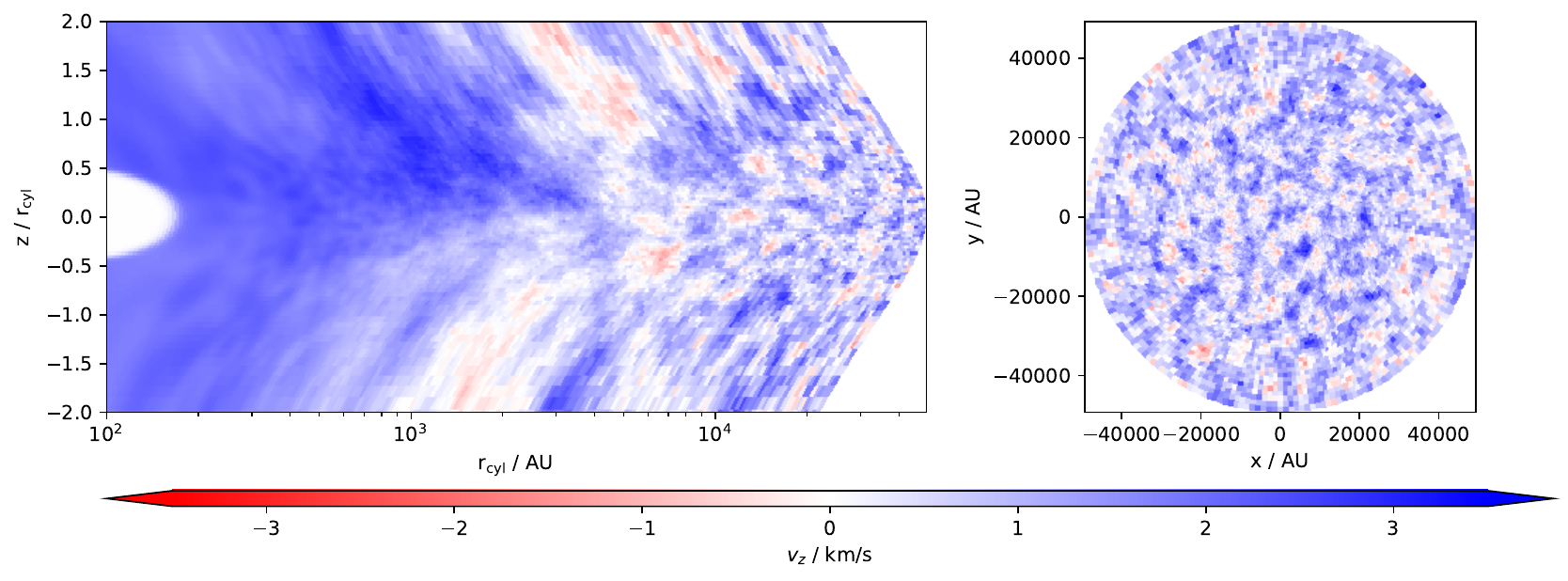}\\
    \includegraphics[width=\linewidth]{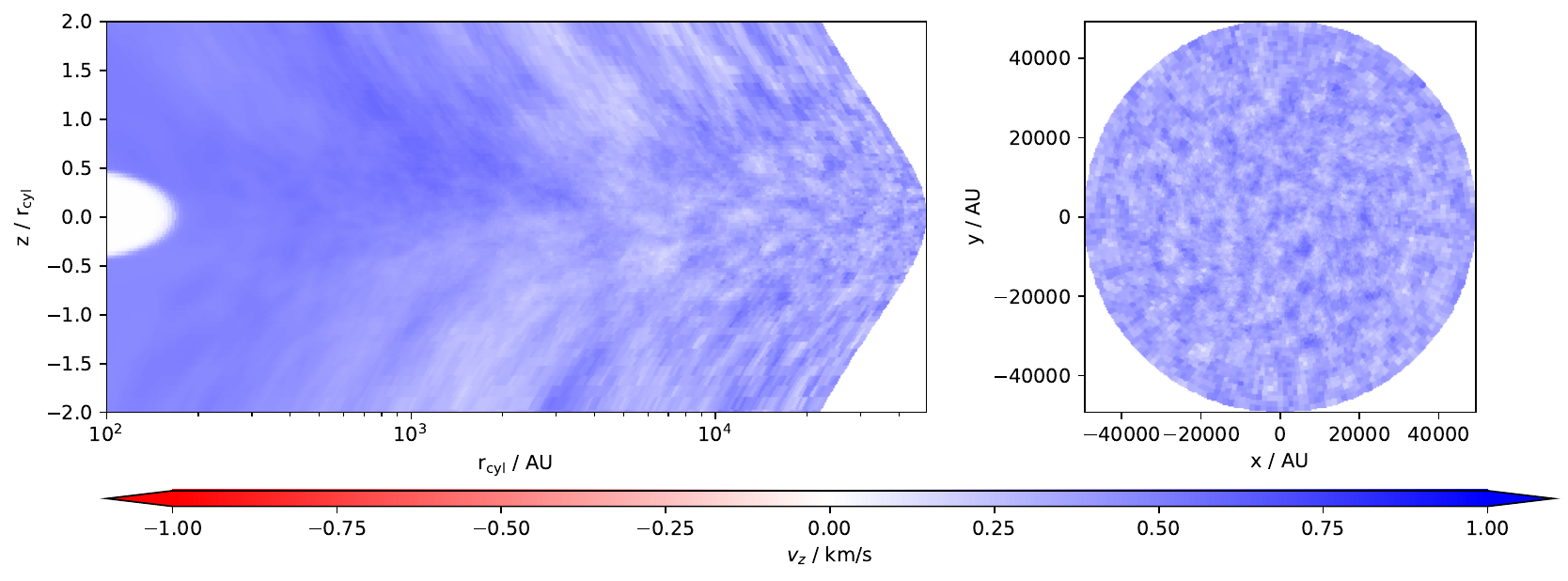}
    \caption{Velocity in Cartesian $z$ direction, $v_z$, at $t=0$ of simulations number 1 and 5. The left column panels show a cylindrical slice at $\phi=0$, and the right column panels show a Cartesian slice at the simulation midplane, $\theta=\pi/2$. There are different color scales for the two simulations.}
    \label{fig:b_turbfield}
\end{figure*}%
\noindent We modeled turbulence as a Gaussian random field, whose strength is defined by $\sigma_\mathrm{turb}$. Figure \ref{fig:b_turbfield} shows the realization of this field for one velocity component, Cartesian $v_z$, for simulations number 1 and 5 from Table \ref{tab:bh_params}, which use $\sigma_\mathrm{turb}=\SI{1}{\kilo\meter\per\second}$ and $\sigma_\mathrm{turb}=\SI{0.1}{\kilo\meter\per\second}$, respectively. The resulting turbulent velocity fluctuations are less significant for simulation number 5, but not negligible. We note that the depicted velocity component contains the contribution from the systemic velocity of the disk.
\FloatBarrier\newpage

\section{Low accretion rate}\label{app:low_acc}
\begin{figure}[htp]
    \centering\includegraphics[width=.9\linewidth]{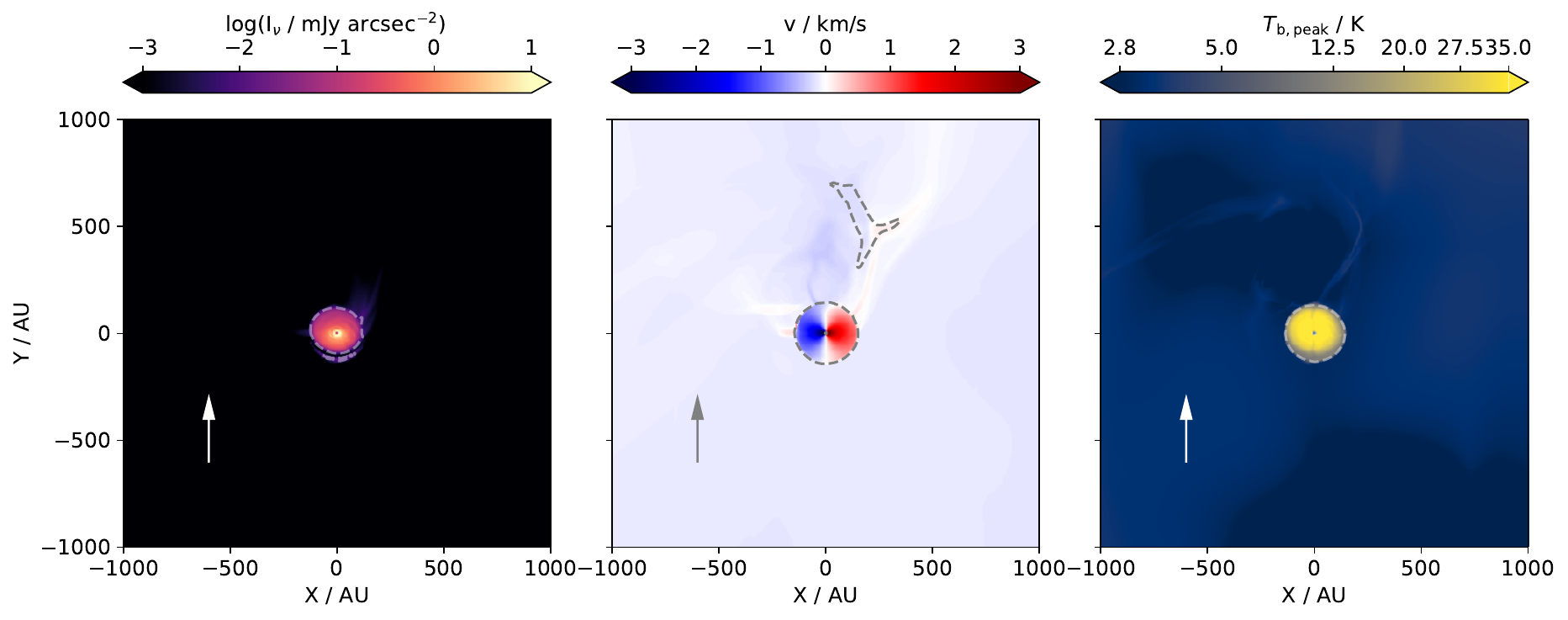}
    \caption{Same as Fig. \ref{fig:b_1_1_d2}, but for simulation number 6, with the level of the moment 0 contour line being \SI{60}{\milli\jansky\per\arcsecond\squared\kilo\meter\per\second}.}
    \label{fig:b_0.5_0.5_d2_9}
\end{figure}%
The fallback streamers discussed in the main text emerge in a dense environment, that is, where $\dot{M}_\mathrm{sys}=\SI{e-8}{\solarmass\per\year}$. However, some stars may undergo considerably lower accretion rates, down to ${\sim}\SI{e-9}{\solarmass\per\year}$ (e.g., \citealt{herczeg2023}). If less material is accreted onto the disk, related accretion signatures are fainter, and streamers may become undetectable. Therefore, we investigated a low accretion rate of \SI{e-9}{\solarmass\per\year} in simulation number 6, with a setup that is otherwise identical to simulation number 2, where multiple fallback streamers are visible.

We find that, while the emerging structures are qualitatively similar dynamically, the streamers can indeed no longer be detected, as shown in Fig. \ref{fig:b_0.5_0.5_d2_9}. In the peak brightness temperature map, only a single component of the streamer originating from the top right can be seen; all other streamers and their subcomponents can no longer be detected. Its emission is insignificant ($T_\mathrm{b,peak}<\SI{5}{\kelvin}$) compared to the emission in all previous cases and compared to the disk emission, so that this system would not be detected to be moving through an environment with strong turbulence. Almost no scattered light emission can be seen at flux levels ${>}\SI{e-3}{\milli\jansky\per\arcsecond\squared}$, and only a small region of the top right streamer can be seen in the moment 0 map with a faint signal of $\SI{60}{\milli\jansky\per\arcsecond\squared\kilo\meter\per\second}$. We therefore conclude that, in order for fallback streamers to be a plausible formation scenario for observed streamers, the infall rate needs to be of the order of ${\sim}\SI{e-8}{\solarmass\per\year}$.
\FloatBarrier\newpage

\section{Single velocity channels}\label{app:channels}
\begin{figure*}[htp]
    \centering\includegraphics[width=\linewidth]{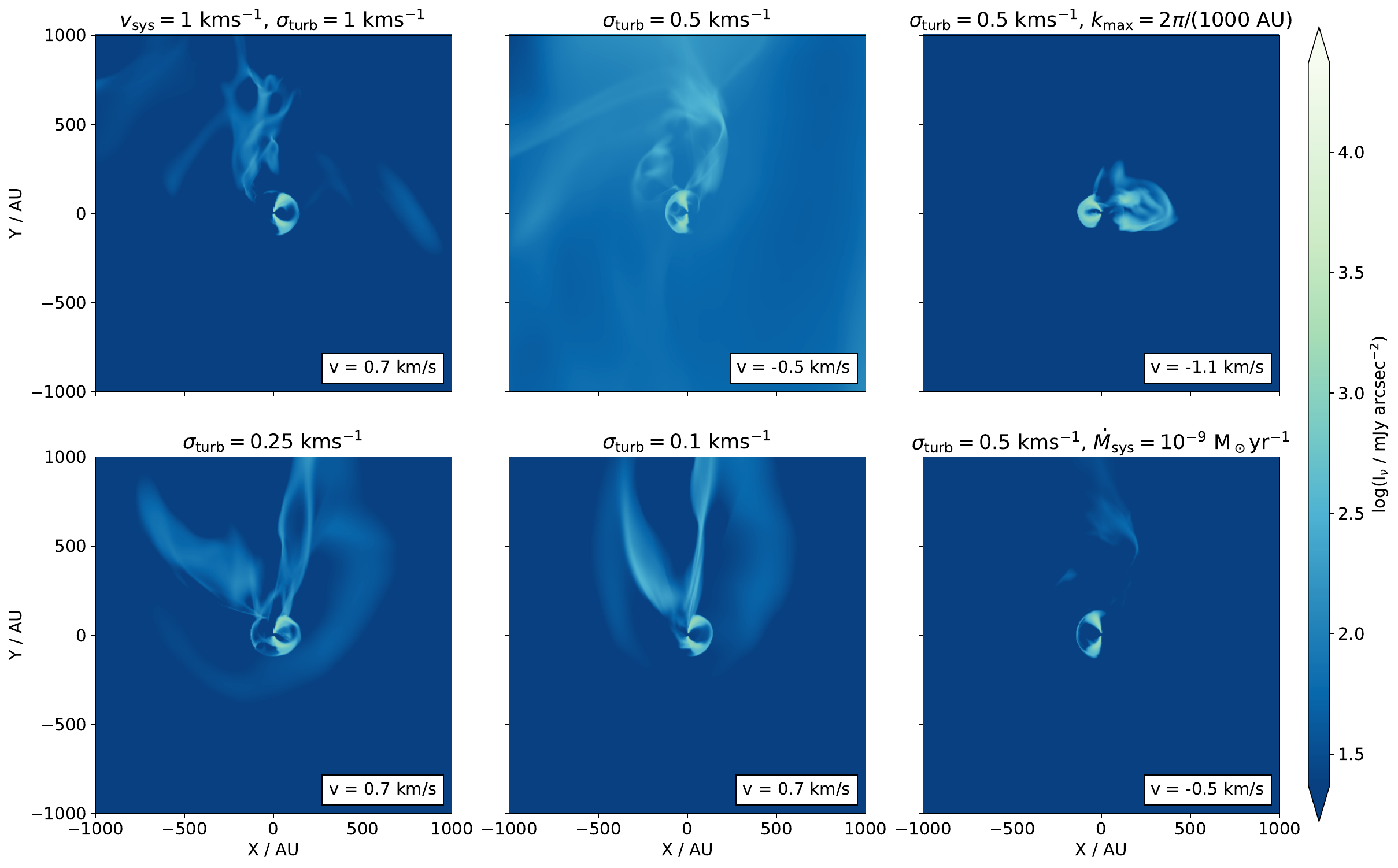}
    \caption{Intensity of the CO emission in a single velocity channel for all simulations listed in Table \ref{tab:bh_params}, at the points in time discussed in Section \ref{sec:res_bhl}. The single channel intensity is computed by averaging five consecutive points in frequency space from the \texttt{RADMC3D} simulation, resulting in a channel width of $\Delta v=\SI{0.2}{\kilo\meter\per\second}$. Unless stated otherwise, the shown simulations have $v_\mathrm{sys}=\SI{0.5}{\kilo\meter\per\second}$ and $k_\mathrm{min}=2\pi/\SI{50}{\astronomicalunit}$.}
    \label{fig:b_channels}
\end{figure*}%
\noindent Figure \ref{fig:b_channels} shows a summary of the CO emission of the various streamers we find in this work. The emission in a single velocity channel is shown, chosen so that key features of the morphology are visible.
\end{appendix}
\end{document}